\documentclass[article]{revtex4}
\usepackage[]{natbib}
\usepackage{graphicx}
\usepackage{fancyhdr}
\usepackage{epsfig}
\RequirePackage{xspace}
\usepackage{relsize}

\def\babar{\mbox{\slshape B\kern-0.1em{\smaller A}\kern-0.1em  B\kern-0.1em{\smaller A\kern-0.2em R}}}

\def\mmu        {\ensuremath{\mu}\xspace}
\def\mtau       {\ensuremath{\tau}\xspace}
\def\taumg      {\ensuremath{\mtau^{\pm} \to \mmu^{\pm} \g}\xspace}
\def\taueg      {\ensuremath{\mtau^{\pm} \to e^{\pm} \g}\xspace}
\def\taulg      {\ensuremath{\mtau^{\pm} \to \ell^{\pm} \g}\xspace}
\def\BRtaueg    {\ensuremath{\BR(\taueg)}\xspace}
\def\BRtaumg    {\ensuremath{\BR(\taumg)}\xspace}

\def\epem       {\ensuremath{e^+e^-}\xspace}
\def\mup        {\ensuremath{\mu^+}\xspace}

\def\taup       {\ensuremath{\tau^+}\xspace}

\def\tautau     {\ensuremath{\tau^+\tau^-}\xspace}

\def\ellp       {\ensuremath{\ell^+}\xspace}

\def\nub        {\ensuremath{\overline{\nu}}\xspace}
\def\nue        {\ensuremath{\nu_e}\xspace}

\def\g     {\ensuremath{\gamma}\xspace}

\def\Wp     {\ensuremath{W^+}\xspace}

\def\qqbar {\ensuremath{q\overline q}\xspace}

\def\ccbar {\ensuremath{c\overline c}\xspace}

\def\bbbar {\ensuremath{b\overline b}\xspace}

\def\piz   {\ensuremath{\pi^0}\xspace}

\def\pip   {\ensuremath{\pi^+}\xspace}
\def\pim   {\ensuremath{\pi^-}\xspace}

\def\pipm  {\ensuremath{\pi^\pm}\xspace}

\def\Kbar  {\kern 0.2em\overline{\kern -0.2em K}{}\xspace}

\def\Kz    {\ensuremath{K^0}\xspace}
\def\Kzb   {\ensuremath{\Kbar^0}\xspace}
\def\KzKzb {\ensuremath{\Kz \kern -0.16em \Kzb}\xspace}
\def\Kp    {\ensuremath{K^+}\xspace}
\def\Km    {\ensuremath{K^-}\xspace}

\def\KpKm  {\ensuremath{\Kp \kern -0.16em \Km}\xspace}
\def\KS    {\ensuremath{K^0_{\scriptscriptstyle S}}\xspace}

\def\Dbar    {\kern 0.2em\overline{\kern -0.2em D}{}\xspace}

\def\Dz      {\ensuremath{D^0}\xspace}
\def\Dzb     {\ensuremath{\Dbar^0}\xspace}
\def\DzDzb   {\ensuremath{\Dz {\kern -0.16em \Dzb}}\xspace}
\def\Dp      {\ensuremath{D^+}\xspace}
\def\Dm      {\ensuremath{D^-}\xspace}

\def\DpDm    {\ensuremath{\Dp {\kern -0.16em \Dm}}\xspace}

\def\Dstarp  {\ensuremath{D^{*+}}\xspace}

\def\Ds      {\ensuremath{D^+_s}\xspace}

\def\Dss     {\ensuremath{D^{*+}_s}\xspace}

\def\Bbar    {\kern 0.18em\overline{\kern -0.18em B}{}\xspace}

\def\Bz      {\ensuremath{B^0}\xspace}
\def\Bzb     {\ensuremath{\Bbar^0}\xspace}
\def\BzBzb   {\ensuremath{\Bz {\kern -0.16em \Bzb}}\xspace}
\def\Bu      {\ensuremath{B^+}\xspace}
\def\Bub     {\ensuremath{B^-}\xspace}

\def\BpBm    {\ensuremath{\Bu {\kern -0.16em \Bub}}\xspace}

\def\BorBbar    {\kern 0.18em\optbar{\kern -0.18em B}{}\xspace}
\def\DorDbar    {\kern 0.18em\optbar{\kern -0.18em D}{}\xspace}
\def\KorKbar    {\kern 0.18em\optbar{\kern -0.18em K}{}\xspace}

\mathchardef\Upsilon="7107
\def\Y#1S{\ensuremath{\Upsilon{(#1S)}}\xspace}

\def\TwoS  {\Y2S}
\def\ThreeS{\Y3S}
\def\FourS {\Y4S}

\mathchardef\Deltares="7101
\mathchardef\Xi="7104
\mathchardef\Lambda="7103
\mathchardef\Sigma="7106
\mathchardef\Omega="710A

\def\Deltabar{\kern 0.25em\overline{\kern -0.25em \Deltares}{}\xspace}
\def\Lbar{\kern 0.2em\overline{\kern -0.2em\Lambda\kern 0.05em}\kern-0.05em{}\xspace}
\def\Sigbar{\kern 0.2em\overline{\kern -0.2em \Sigma}{}\xspace}
\def\Xibar{\kern 0.2em\overline{\kern -0.2em \Xi}{}\xspace}
\def\Obar{\kern 0.2em\overline{\kern -0.2em \Omega}{}\xspace}
\def\Nbar{\kern 0.2em\overline{\kern -0.2em N}{}\xspace}
\def\Xb{\kern 0.2em\overline{\kern -0.2em X}{}\xspace}

\def\BR         {{\ensuremath{\cal B}\xspace}}

\def\Btopilnu   {\ensuremath{B \to \pi \ell \nu}\xspace}

\def\mec        {\mbox{$m_{\rm EC}$}\xspace}
\def\DeltaE     {\mbox{$\Delta E$}\xspace}

\newcommand{\tev}{\ensuremath{\mathrm{\,Te\kern -0.1em V}}\xspace}
\newcommand{\gev}{\ensuremath{\mathrm{\,Ge\kern -0.1em V}}\xspace}
\newcommand{\mev}{\ensuremath{\mathrm{\,Me\kern -0.1em V}}\xspace}
\newcommand{\kev}{\ensuremath{\mathrm{\,ke\kern -0.1em V}}\xspace}
\newcommand{\ev}{\ensuremath{\mathrm{\,e\kern -0.1em V}}\xspace}
\newcommand{\gevc}{\ensuremath{{\mathrm{\,Ge\kern -0.1em V\!/}c}}\xspace}
\newcommand{\mevc}{\ensuremath{{\mathrm{\,Me\kern -0.1em V\!/}c}}\xspace}
\newcommand{\gevcc}{\ensuremath{{\mathrm{\,Ge\kern -0.1em V\!/}c^2}}\xspace}
\newcommand{\mevcc}{\ensuremath{{\mathrm{\,Me\kern -0.1em V\!/}c^2}}\xspace}

\def\invfb   {\ensuremath{\mbox{\,fb}^{-1}}\xspace}

\def\mus  {\ensuremath{\rm \,\mus}\xspace}

\def\mus        {\ensuremath{\,\mu{\rm s}}\xspace}    

\def\ra                 {\ensuremath{\rightarrow}\xspace}
\def\to                 {\ensuremath{\rightarrow}\xspace}

\newcommand{\stat}{\ensuremath{\mathrm{(stat)}}\xspace}
\newcommand{\syst}{\ensuremath{\mathrm{(syst)}}\xspace}

\def\pep2{PEP-II}

\def\gsim{{~\raise.15em\hbox{$>$}\kern-.85em
          \lower.35em\hbox{$\sim$}~}\xspace}
\def\lsim{{~\raise.15em\hbox{$<$}\kern-.85em
          \lower.35em\hbox{$\sim$}~}\xspace}

\def\CP                {\ensuremath{C\!P}\xspace}
\def\CPT               {\ensuremath{C\!PT}\xspace} 

\def\Vcs  {\ensuremath{|V_{cs}|}\xspace}

\def\Vub  {\ensuremath{|V_{ub}|}\xspace}

\xspace

\def\etal{{\it et\,al.}} 

\def\Bzpilnu{\ensuremath{B^{0} \rightarrow \pi^-\ell^+\nu}\xspace}
\def\Bppizlnu{\ensuremath{B^{+} \rightarrow \pi^0\ell^+\nu}\xspace}
\def\Bzrholnu{\ensuremath{B^{0} \rightarrow \rho^-\ell^+\nu}\xspace}
\def\Bprhozlnu{\ensuremath{B^{+} \rightarrow \rho^0\ell^+\nu}\xspace}

\def\bulnu{\ensuremath{B \rightarrow X_u\ell\nu}\xspace}

\def\mES{\ensuremath{m_{\rm ES}}\xspace}
\def\DeltaE{\ensuremath{\Delta E}\xspace}

\def\invfb    {\ensuremath{\mbox{\,fb}^{-1}}\xspace}
\def\Ds       {\ensuremath{D^+_s}\xspace}
\def\Dz       {\ensuremath{D^0}\xspace}
\def\Dbar     {\kern 0.2em\overline{\kern -0.2em D}{}\xspace}
\def\Dzb      {\ensuremath{\Dbar^0}\xspace}
\def\Dp       {\ensuremath{D^+}\xspace}
\def\Dm       {\ensuremath{D^-}\xspace}

\def\taup     {\ensuremath{\tau^+}\xspace}
\def\nut      {\ensuremath{\nu_\tau}\xspace}
\def\nutb     {\ensuremath{\nub_\tau}\xspace}

\def\ep       {\ensuremath{e^+}\xspace}
\def\en       {\ensuremath{e^-}\xspace}  
\def\epem     {\ensuremath{e^+e^-}\xspace}
\def\nue      {\ensuremath{\nu_e}\xspace}
\def\nub      {\ensuremath{\overline{\nu}}\xspace}

\def\ccbar    {\ensuremath{c\overline c}\xspace}
\def\Bbar     {\kern 0.18em\overline{\kern -0.18em B}{}\xspace}

\def\Wp       {\ensuremath{W^+}\xspace}

\def\ellp     {\ensuremath{\ell^+}\xspace}
\def\ellbar   {\kern 0.18em\overline{\kern -0.18em \ell}{}\xspace}
\def\Vcs      {\ensuremath{|V_{cs}|}\xspace}

\def\mup      {\ensuremath{\mu^+}\xspace}
\def\Dss      {\ensuremath{D^{*+}_s}\xspace}
\def\piz      {\ensuremath{\pi^0}\xspace}
\def\pip      {\ensuremath{\pi^+}\xspace}
\def\pim      {\ensuremath{\pi^-}\xspace}
\def\pipm     {\ensuremath{\pi^\pm}\xspace}
 
\def\KS       {\ensuremath{K^0_{\scriptscriptstyle S}}\xspace} 
\def\Kp       {\ensuremath{K^+}\xspace}
\def\Km       {\ensuremath{K^-}\xspace}

\def\Kbar     {\kern 0.2em\overline{\kern -0.2em K}{}\xspace}
\def\BR       {{\ensuremath{\cal B}\xspace}}
\def\pep2     {{PEP-II~}}
\def\g        {\ensuremath{\gamma}\xspace}
\def\qqbar    {\ensuremath{q\overline q}\xspace}

\def\fDs      {\ensuremath{f_{D_s}}\xspace}

\def\pisoftp    {\ensuremath{\pi_{\rm s}^{+}}\xspace}

\def\Kmpip      {\ensuremath{K^{-}\pi^{+}}\xspace}
\def\KmKp       {\ensuremath{K^{-}K^{+}}\xspace}
\def\KpKm       {\ensuremath{K^{+}K^{-}}\xspace}

\newcommand{\kevcc}{\ensuremath{{\mathrm{\,Ke\kern -0.1em V\!/}c^2}}\xspace}
\def\dm         {\ensuremath{\Delta m}\xspace}

\catcode`\@=11\relax
\newskip\dkwidth
\def\dk{%
   \dkwidth=2em plus 0.5 em minus 0.25 em\relax
   {\m@th\mathord{%
   \hbox{%
      \kern 0.3em
      \raise 0.6ex%
      \hbox{%
         \vrule width 0.25pt height 0.5\dkwidth depth0pt}%
      \kern-1.2pt%
      \hbox to 1.1\dkwidth{%
         \rightarrowfill}%
      \kern0.4em}}%
   }%
}%
\def\rightarrowfill{$\m@th\mathord-\mkern-10mu%
  \cleaders\hbox{$\mkern-2mu\mathord-\mkern-2mu$}\hfill
  \mkern-6mu\mathord\rightarrow$}
\catcode`\@=12\relax
\def\yCP        {\ensuremath{y_{C\!P}}\xspace}

\def\Kmpip      {\ensuremath{\Km\pip}\xspace}

\def\tauKpi     {\ensuremath{\tau_{K\pi}}\xspace}
\def\tauhh      {\ensuremath{\tau_{hh}}\xspace}

\def\taup       {\ensuremath{\tau^+}\xspace}

\def\to                 {\ensuremath{\rightarrow}\xspace}
\def\pisoftp    {\ensuremath{\pi_{\rm s}^{+}}\xspace}
\def\mKKpipi {\ensuremath{m(\Kp\Km\pip\pim)}\xspace}
\def\dm         {\ensuremath{\Delta m}\xspace}
\mathchardef\Upsilon="7107
\def\Y#1S{\ensuremath{\Upsilon{(#1S)}}\xspace}
\def\FourS {\Y4S}
\def\Dbar    {\kern 0.2em\overline{\kern -0.2em D}{}\xspace}

\def\Dstarp  {\ensuremath{D^{*+}}\xspace}
\def\Ds      {\ensuremath{D^+_s}\xspace}
\def\Dz      {\ensuremath{D^0}\xspace}
\def\Dzb     {\ensuremath{\Dbar^0}\xspace}
\def\Kp    {\ensuremath{K^+}\xspace}
\def\Km    {\ensuremath{K^-}\xspace}
\def\KS    {\ensuremath{K^0_{\scriptscriptstyle S}}\xspace}
\def\pip   {\ensuremath{\pi^+}\xspace}
\def\pim   {\ensuremath{\pi^-}\xspace}
\def\sys {\hbox{syst}}
\def\sta {\hbox{stat}}
\def\pep2{PEP-II}
\def\invfb   {\ensuremath{\mbox{\,fb}^{-1}}\xspace}
\def\epem       {\ensuremath{e^+e^-}\xspace}
\def\ccbar {\ensuremath{c\overline c}\xspace}
\def\CP                {\ensuremath{C\!P}\xspace}
\def\CPT               {\ensuremath{C\!PT}\xspace} 

\def\DeltaE                    {\ensuremath{\Delta E}\xspace}

\def\eff                       {\ensuremath{\varepsilon}\xspace}

\newcommand{\roots}        {\ensuremath{\sqrt{s}}\xspace}

\newcommand{\gevccgevcc}{\ensuremath{{\mathrm{\,Ge\kern -0.1em V^2\!/}c^4}}\xspace}
\newcommand{\evcc}{\ensuremath{{\mathrm{\,e\kern -0.1em V\!/}c^2}}\xspace}
\newcommand{\CM} {\mbox{CM}\xspace}

\begin{document}

\title{Recent Results on Flavor Physics from \babar}

%

\author{J. Benitez on behalf of the \babar~ Collaboration}
\affiliation{SLAC, Stanford, CA 94025, USA}

\begin{abstract}
We report an update to our previous measurement of the CKM element $\Vub$ using exclusive \Btopilnu decays. In the charm sector we have performed a measurement of \fDs using $\Ds\rightarrow\taup\nut$ decays, we have measured the mixing parameter \yCP using the lifetime ratio $\frac{\langle\tauKpi\rangle}{\langle\tauhh\rangle}$ in \Dz decays, and we have also searched for CP violation using T-odd correlations in  \Dz decays to $\Kp\Km\pip\pim$. Finally, in the tau sector we have performed a search for the lepton flavor violating decays $\tau^\pm \rightarrow e^\pm \gamma$ and $\tau^\pm \rightarrow \mu^\pm \gamma$.
\end{abstract}

\maketitle

\thispagestyle{fancy}


\section{Introduction}
The physics reach of the \babar~ experiment encompasses a large part of the {\it flavor} sector of particle physics. 
The $\babar$ detector at the SLAC \pep2 asymmetric-energy \epem collider collected approximately 500 \invfb at center of mass (CM) energies near 10.58 GeV between 1999 and 2008. At this energy the production cross section for production of fermion-anti-fermion pairs \ccbar, \bbbar and \tautau  is roughly the same and yields an event sample of about 600 Million produced events of each type. These samples enable precise measurements of Cabibbo-Kobayashi-Maskawa (CKM) parameters, studies of charm meson properties, and searches for rare $\tau$ decays among other topics.
This article presents a summary of recent measurements in these areas from \babar.

\section{\large\bf Determination of \Vub from \boldmath \Btopilnu Decays  }

The elements of the CKM quark-mixing matrix are fundamental parameters of the Standard Model (SM) of electroweak interactions. 
With the increasingly precise measurements of decay-time-dependent \CP\ asymmetries in $B$-meson decays, in particular sin(2$\beta$)~\citep{ref:beta}\citep{ref:phi1}, improved measurements of \Vub and $|V_{cb}|$ will allow for more stringent experimental tests of the SM mechanism for \CP\ violation~\citep{ref:sm}.  
The best method to determine $\Vub$ is to measure semileptonic decay rates for  \bulnu\ ($X_u$ refers to hadronic states without charm), which is proportional to $\Vub^2$. 
We have performed a study of four exclusive semileptonic decay modes, \Bzpilnu, \Bppizlnu, \Bzrholnu, and \Bprhozlnu, and a determination of $\Vub$. Here $\ell$ refers to a charged lepton, either \ep\ or \mup , and $\nu$ refers to the associated neutrino. Exclusive decays offer good kinematic constraints, and thus effective background suppression.
This analysis represents an update of an earlier measurement~\citep{ref:pilnu_jochen} that was based on a significantly smaller data set. For the current analysis, the signal yields and background suppression have been improved, and the systematic uncertainties have been reduced through the use of improved reconstruction and signal extraction methods. 

\begin{figure}[ht]    
    \begin{minipage}{0.73\linewidth}
      \epsfig{file=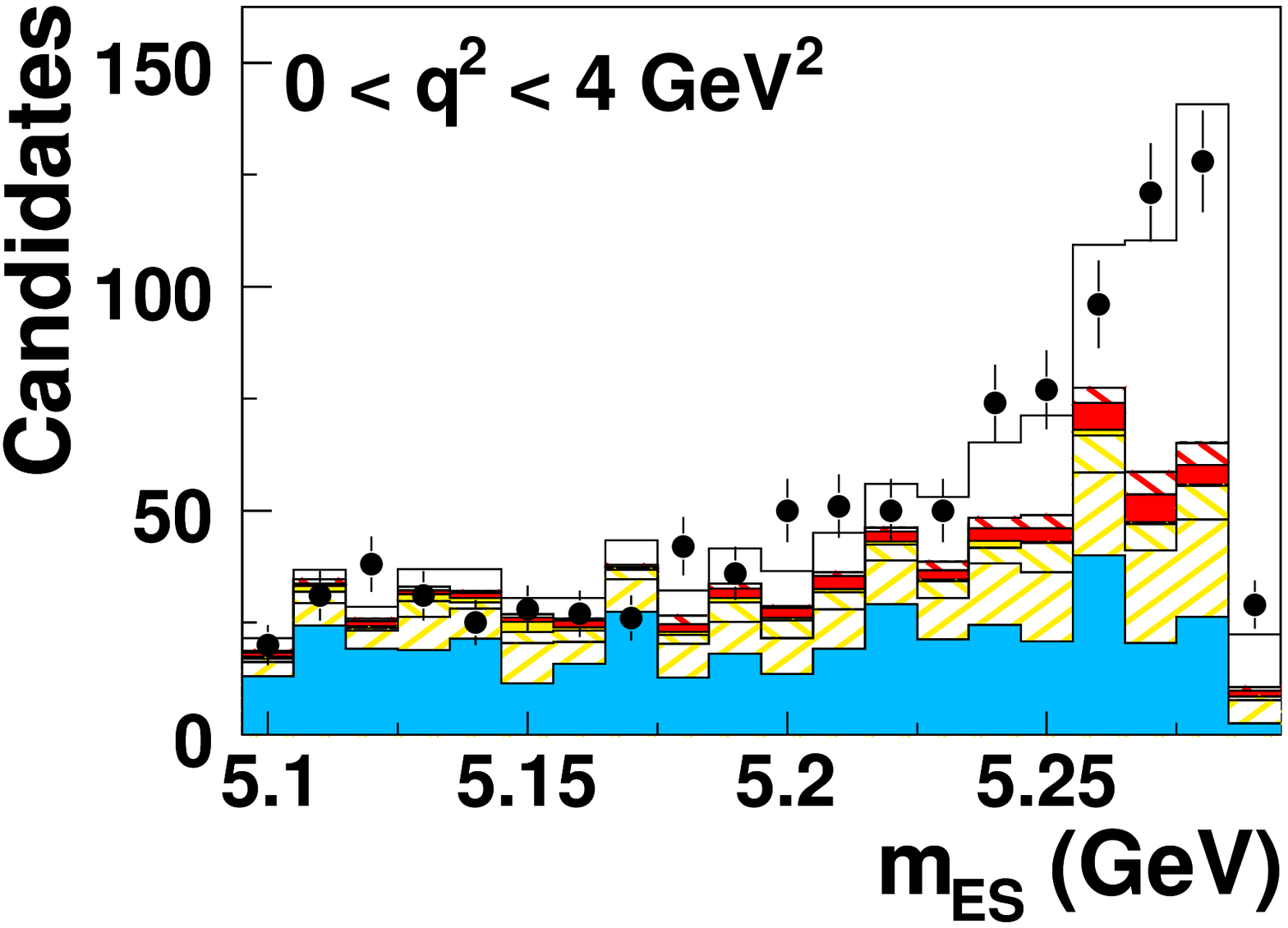,width=.325\linewidth,clip}
      \epsfig{file=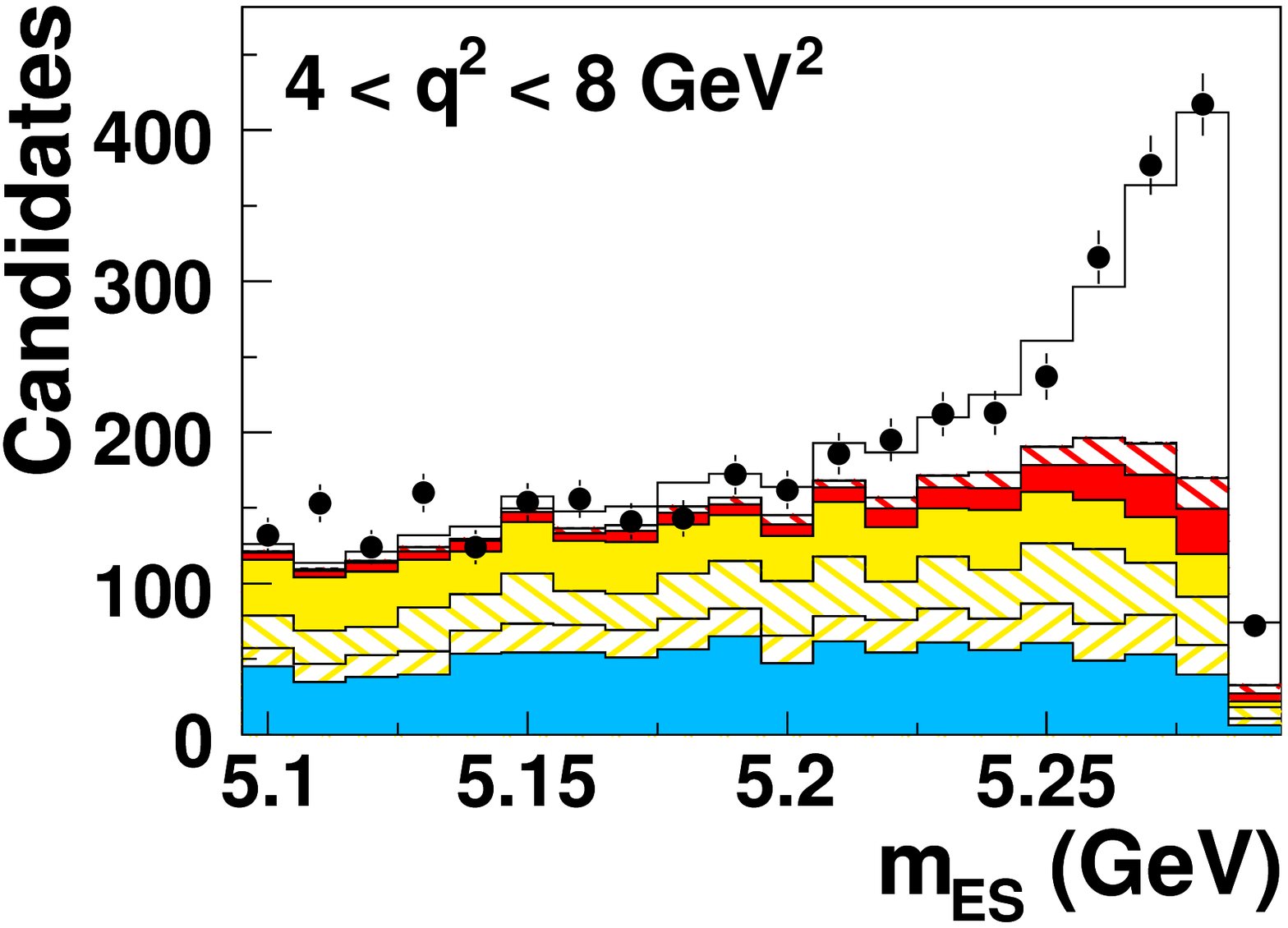,width=.325\linewidth,clip}
      \epsfig{file=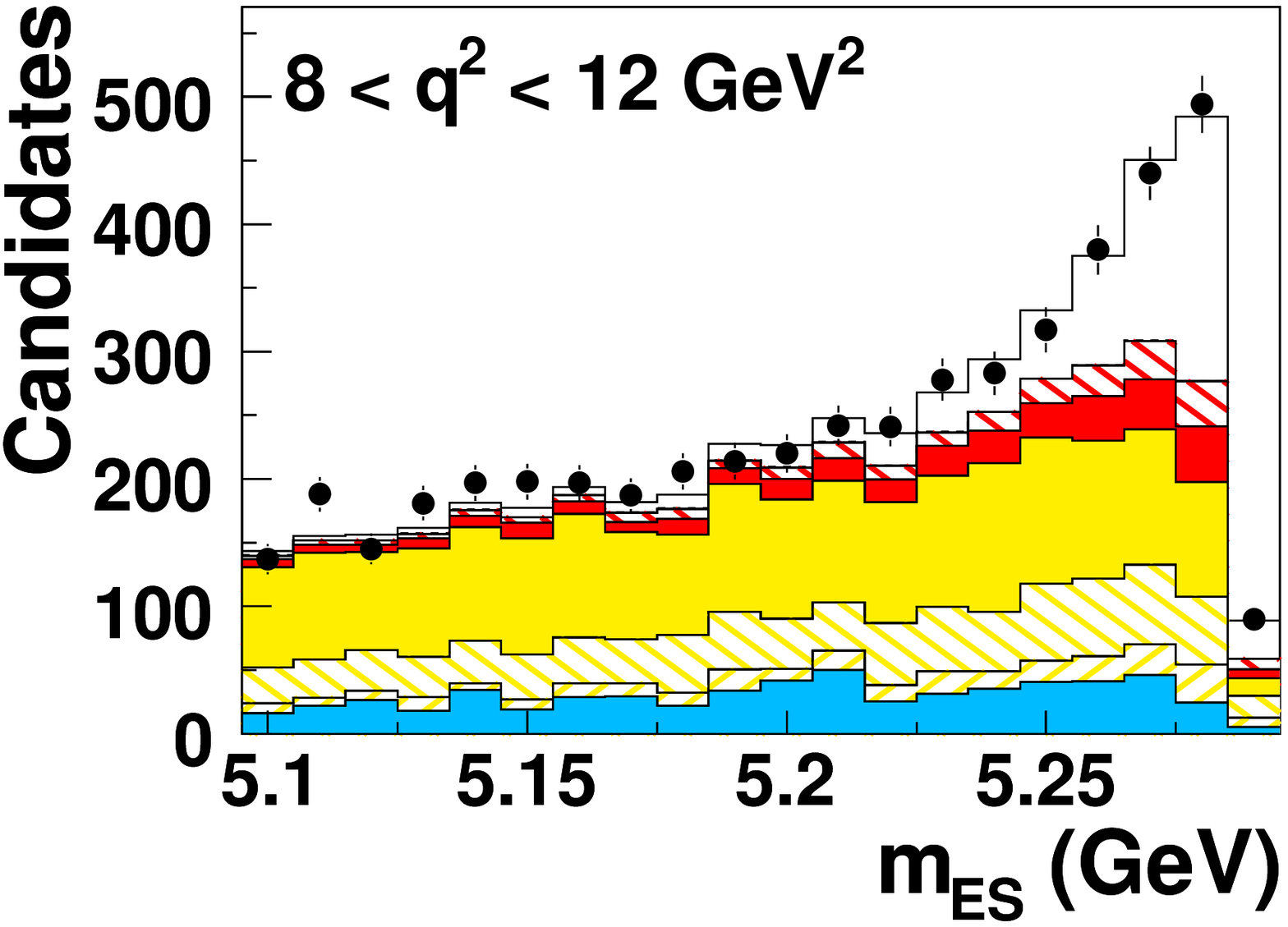,width=.325\linewidth,clip}
      \epsfig{file=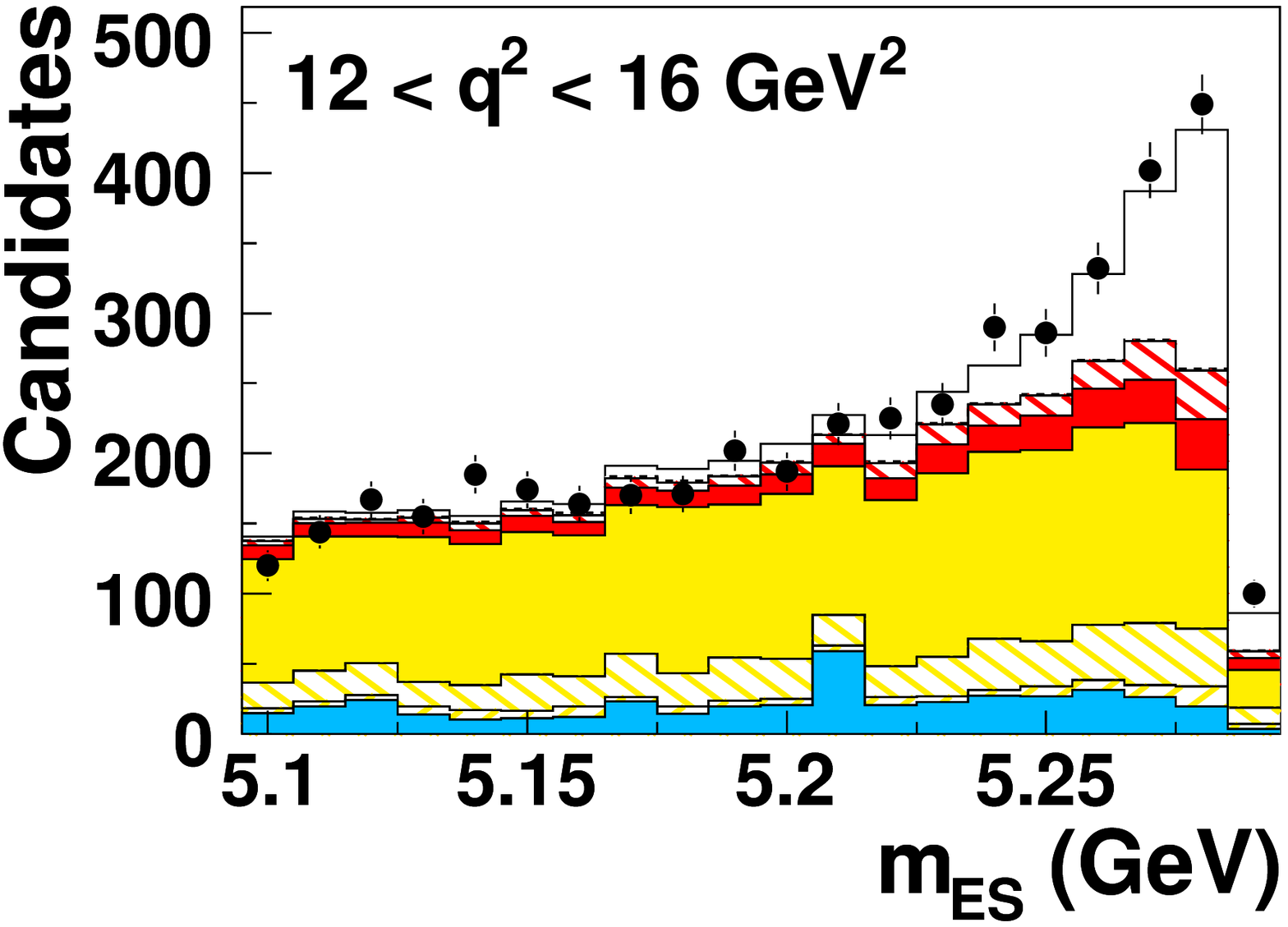,width=.325\linewidth,clip}
      \epsfig{file=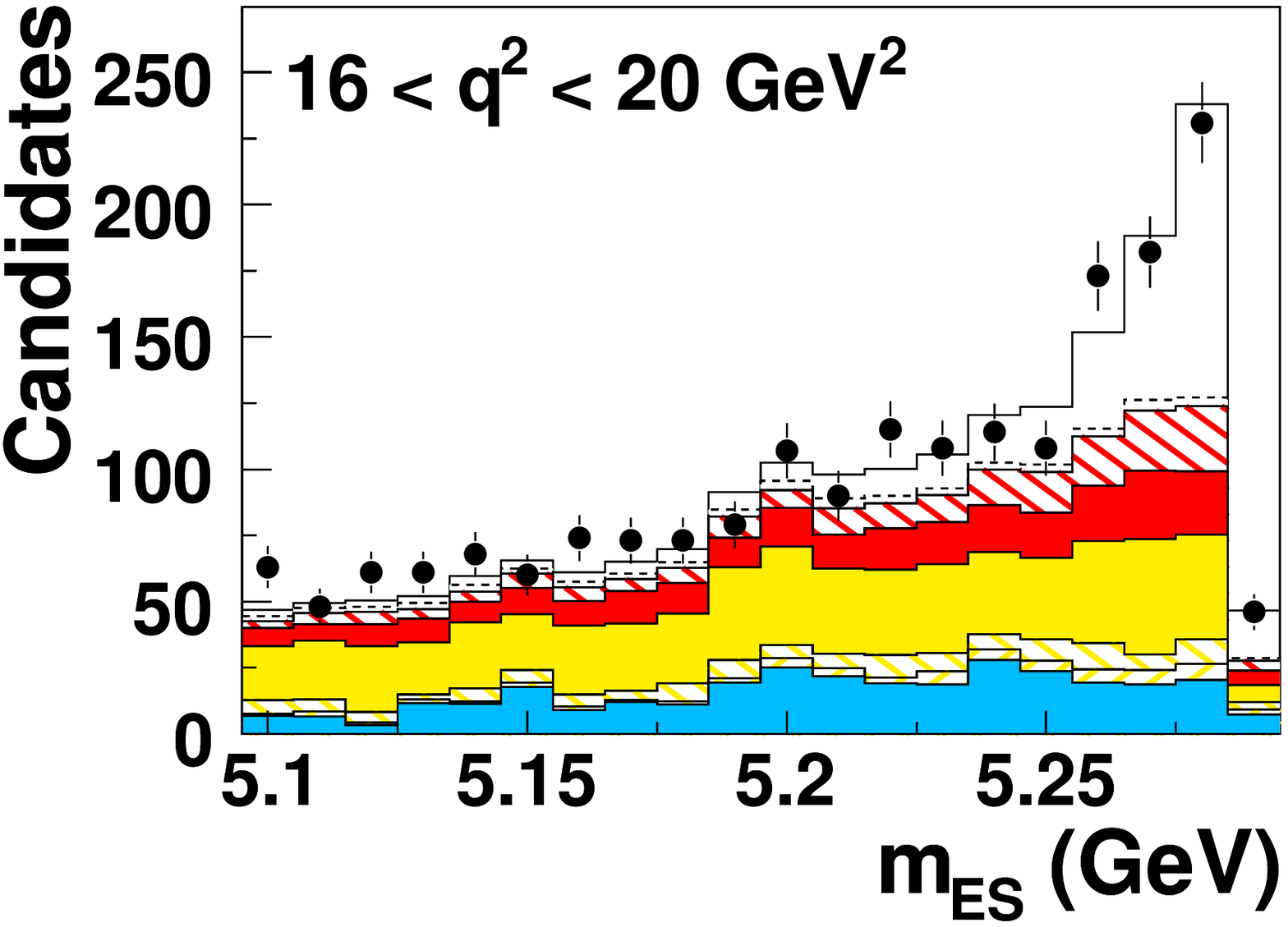,width=.325\linewidth,clip}
      \epsfig{file=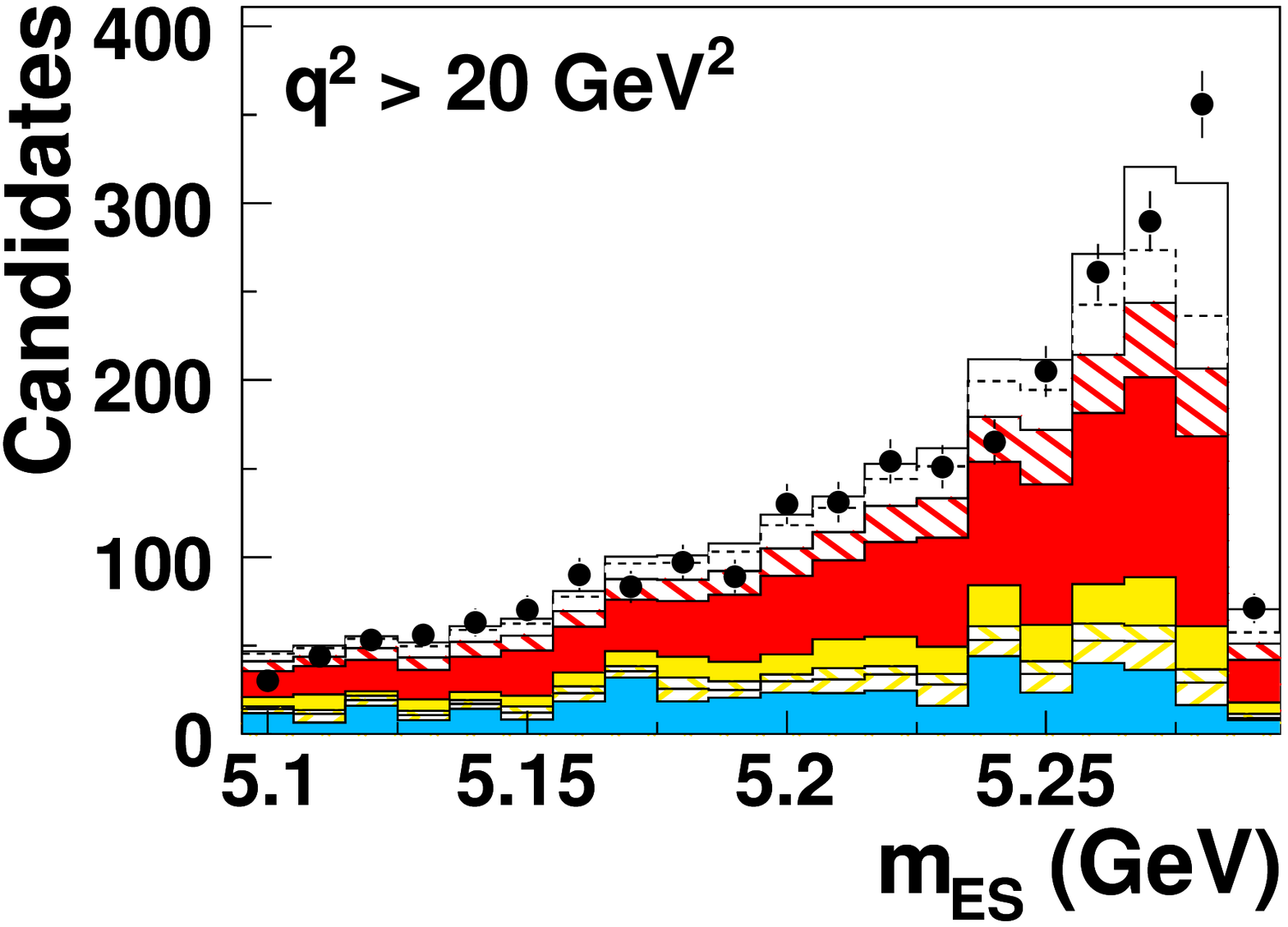,width=.325\linewidth,clip}
    \end{minipage}
    \begin{minipage}{0.25\linewidth}
      \epsfig{file=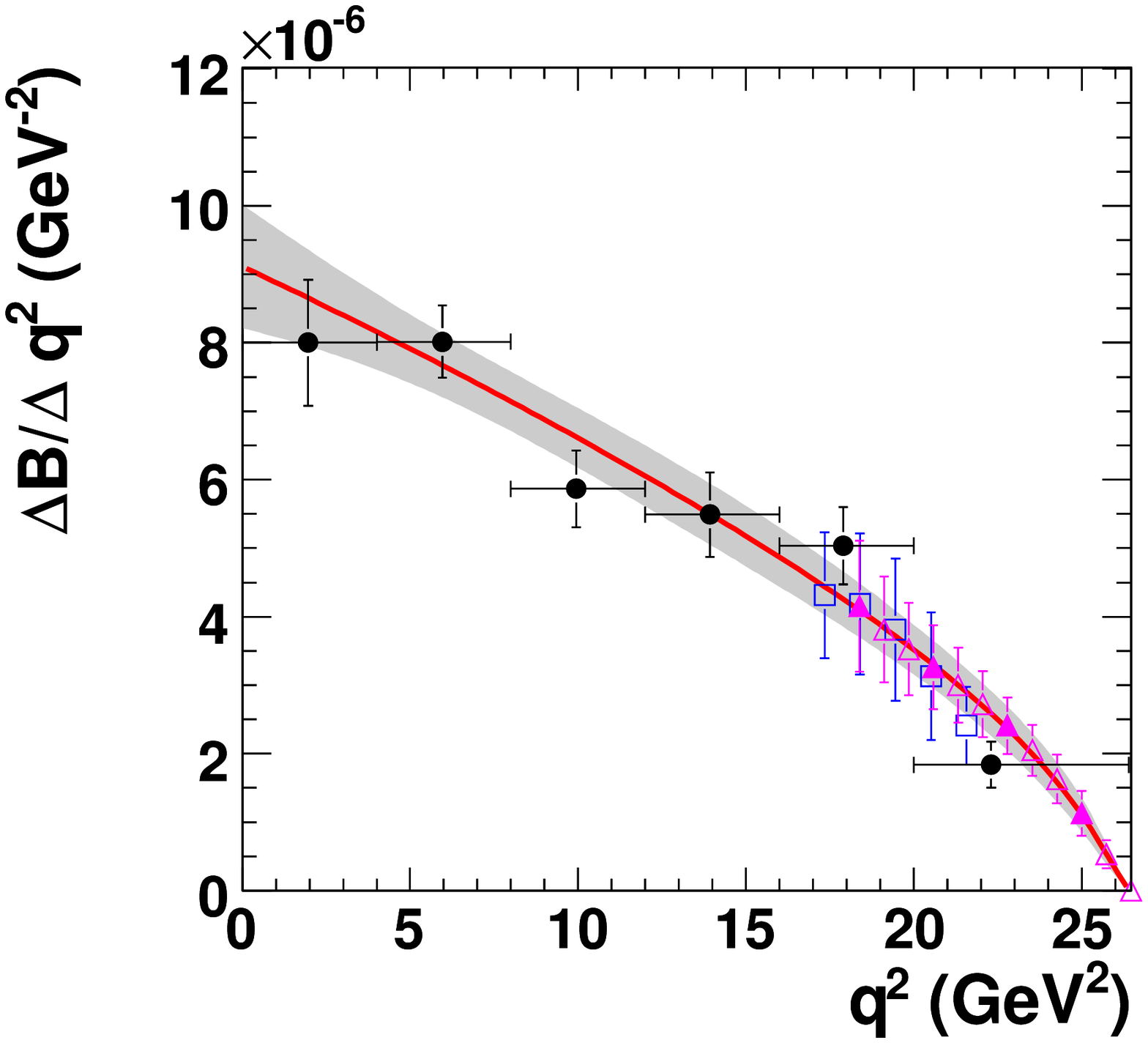,width=\linewidth,clip} 
    \end{minipage}

   \caption{
    Left plots show the \mES  distributions in each $q^2$~bin for \Bzpilnu. The points show the data and the histograms show the fit and the different background contributions.
    The right-most plot shows the simultaneous fit to data (points) and to the FNAL/MILC lattice prediction (magenta, closed triangles). The LQCD results are rescaled according to the \Vub value obtained from the fit.
   }
  \label{fig:DeltaE_pilnu_fit}
\end{figure}

In the analysis of \Bzpilnu we reconstruct the pion and lepton tracks, and determine the neutrino 4-momentum as the missing momentum in the event. A similar reconstruction is performed for the other channels. We then determine the signal yield as a function of $q^2$ by performing a two-dimensional fit in the variables \mES and \DeltaE. \DeltaE is the difference between the reconstructed $B$ energy and half the CM beam energy, and \mES is the mass of the $B$ candidate computed using the reconstructed 3-momentum and half the beam energy. The projections of the fit for \Bzpilnu are shown in Fig.\ref{fig:DeltaE_pilnu_fit}. The fit is performed simultaneously to all channels while constraining \Bppizlnu using isospin symmetry. The \Bzrholnu and \Bprhozlnu modes help constrain the background. From the extracted signal yields the decay rate can be determined using the known total number $B$ events produced and correcting for the reconstruction efficiency.

The decay rate for \Bzpilnu depends on the momentum, $q^2$, carried by the $W^+$ and takes the following formula in the SM,
\begin{eqnarray}
\frac{d\Gamma(B^0 \ra \pi^- \ell^+ \nu)}{dq^2 d \cos\theta_{W\ell}} =
|V_{ub}|^2  \frac{G^2_F \, p_{\pi}^3}
        {32 \pi^3} {\sin}^2\theta_{W\ell} |f_+(q^2)|^2,
\label{semilept:eq:pplnutdd}
\end{eqnarray}  
where $p_\pi$ is the momentum of the pion in the rest frame of the $B$ meson and $\theta_{W\ell}$ is the angle of the charged-lepton momentum in the $W$ rest frame with respect to the direction of the $W$ boost from the $B$ rest frame. The $q^2$-dependent form factor, $f_+(q^2)$, is calculated from Lattice QCD.  Using this formula we convert the form factors into decay rates and require that they match the measured values as shown in Fig.\ref{fig:DeltaE_pilnu_fit}. From the conversion factor \Vub is determined to be $(2.95 \pm 0.31)\times 10^{-3} $.

\section{{\boldmath 
Measurement of the Branching Fraction for $\Ds\rightarrow\taup\nut$ and Extraction of the Decay Constant $f_{D_s}$
}}

The purely leptonic decays of the $\Ds$ meson provide a clean probe of the pseudoscalar meson decay constant $f_{D_s}$, which describes the amplitude for the $c$ and $\bar{s}$ quarks to have zero spatial separation within the meson. In the SM these decays occur through a virtual $\Wp$ boson which decays to a lepton pair, ignoring radiative processes, the total width is
\begin{equation}
  \Gamma(\Ds\rightarrow\ellp\nu_{\ell}) = \frac{G^2_F}{8\pi} M_{\Ds}^3 \left(\frac{m_{\ell}}{M_{\Ds}}\right)^2 \left(1-\frac{m_{\ell}^2}{M_{\Ds}^2}\right)^2\Vcs^2f^2_{D_s},
\label{eq:fDs}
\end{equation}
where $M_{\Ds}$ and $m_{\ell}$ are the $\Ds$ and lepton masses, respectively, $G_F$ is the Fermi coupling constant, $\Vcs$ is the magnitude of the CKM matrix element that characterizes the coupling of the weak charged current to the $c$ and $\bar{s}$ quarks. 
In the context of the SM, predictions for meson decay constants can be obtained from QCD lattice calculations~\citep{ref:qcdsumrules,ref:qlqcd,ref:uqlqcd,ref:fnalmilc,ref:fermimilc}. The most precise theoretical prediction for $f_{D_s}$ is (241$\pm$3) MeV~\citep{ref:uqlqcd}. This value is in slight disagreement with the current measurement of \fDs \citep{ref:CLEO}.
It is important to validate the lattice QCD predictions through measurements of \fDs as these computational methods are also used in other areas such as $B$ meson decays. In addition, it is possible that physics beyond the SM can induce a difference between the theoretical prediction and the measured value.

The relatively large branching fraction for the $\taup$ decay mode motivates the use of the decay sequence $\Ds\rightarrow\taup\nut$, $\taup\rightarrow\ep\nue\nutb$ in this analysis. 
We use the well known branching fraction $\BR(\Ds\rightarrow\KS\Kp)$ for normalization. 
This analysis uses an integrated luminosity of 427 $\invfb$ corresponding to the production of approximately 554 million $\ccbar$ events.

\begin{figure}[ht]
\begin{center}
  \includegraphics[width=.24\linewidth,height=.24\linewidth]{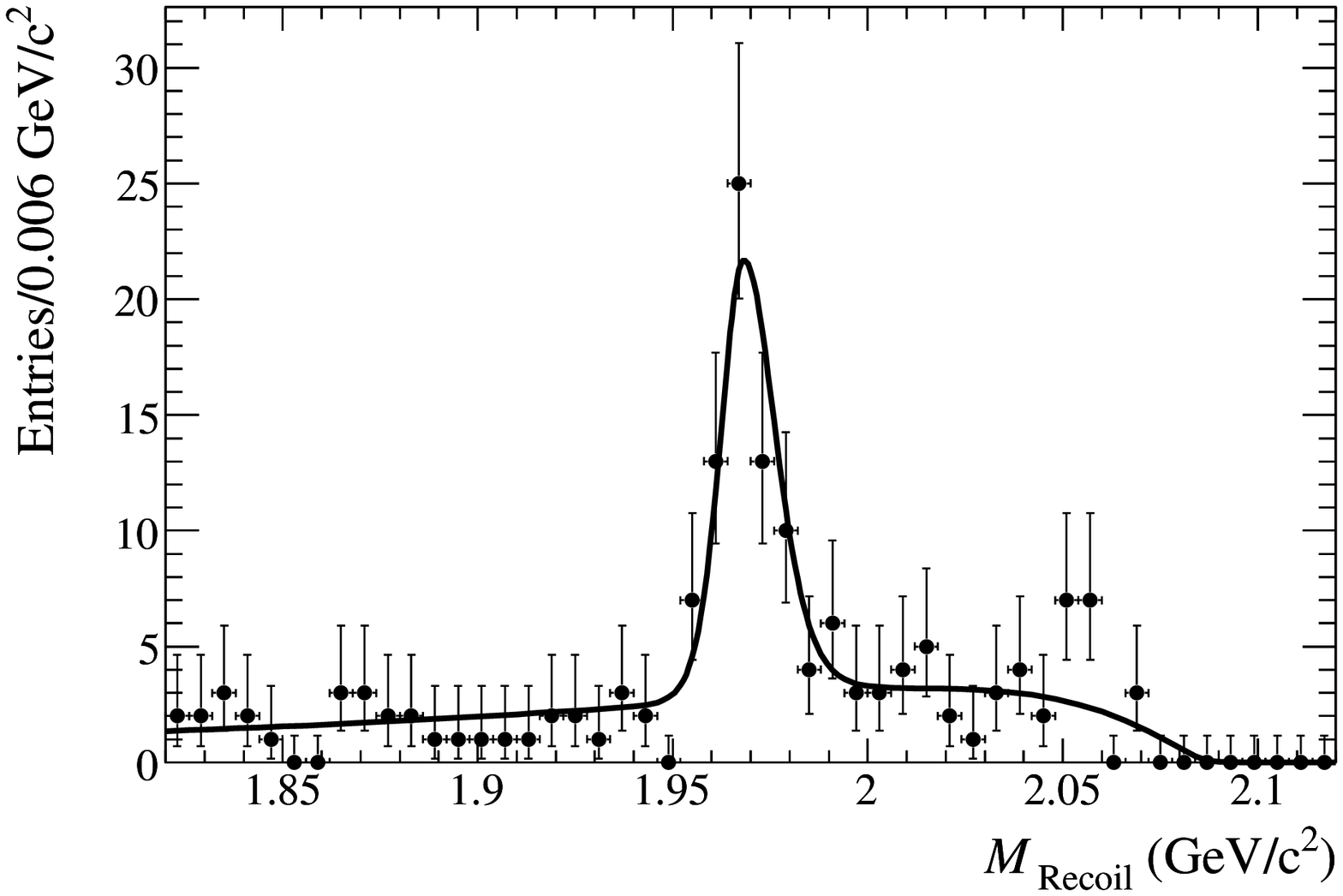} 
  \includegraphics[width=.24\linewidth,height=.24\linewidth]{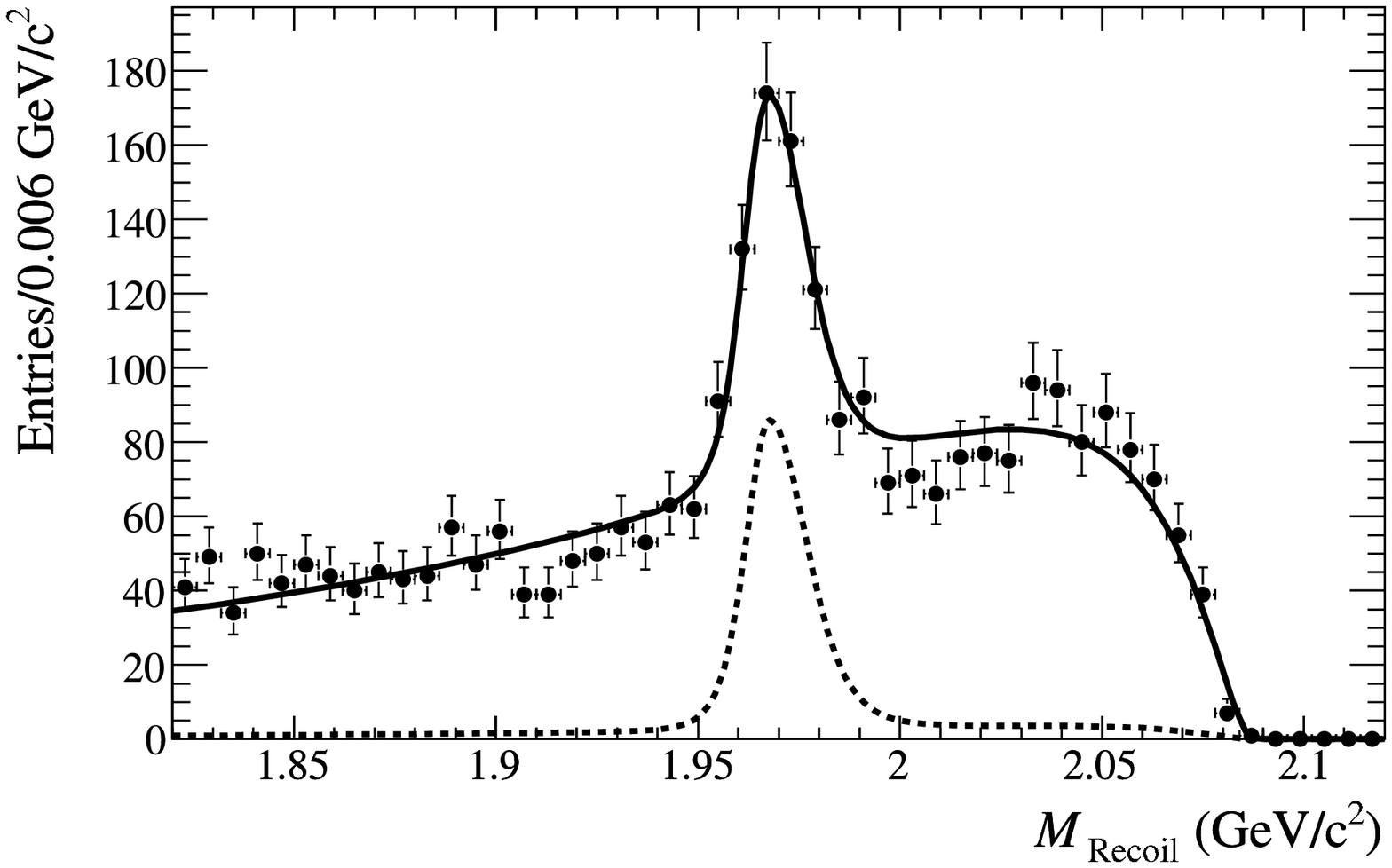} 
  \includegraphics[width=.24\linewidth,height=.24\linewidth]{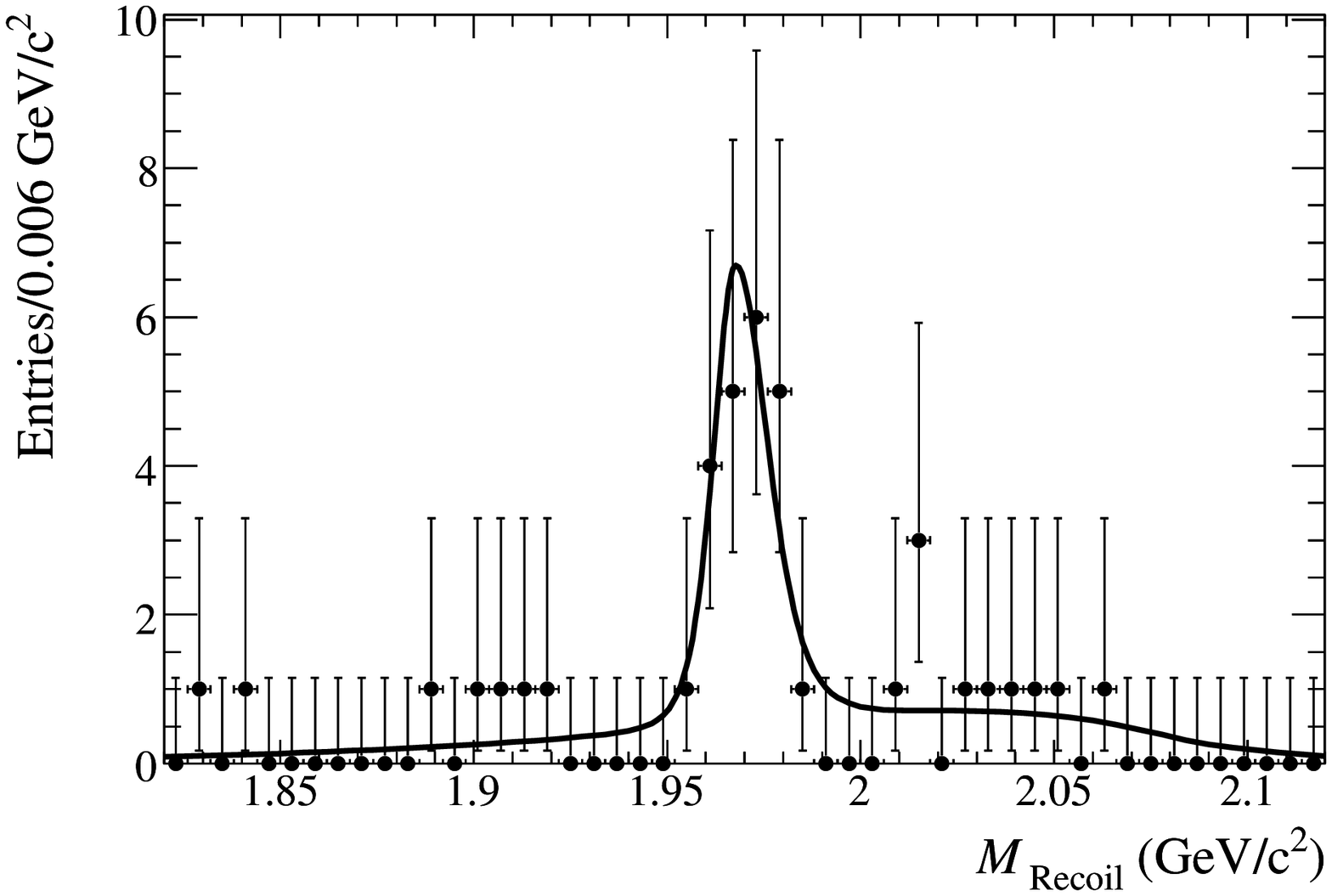} 
  \includegraphics[width=.24\linewidth,height=.24\linewidth]{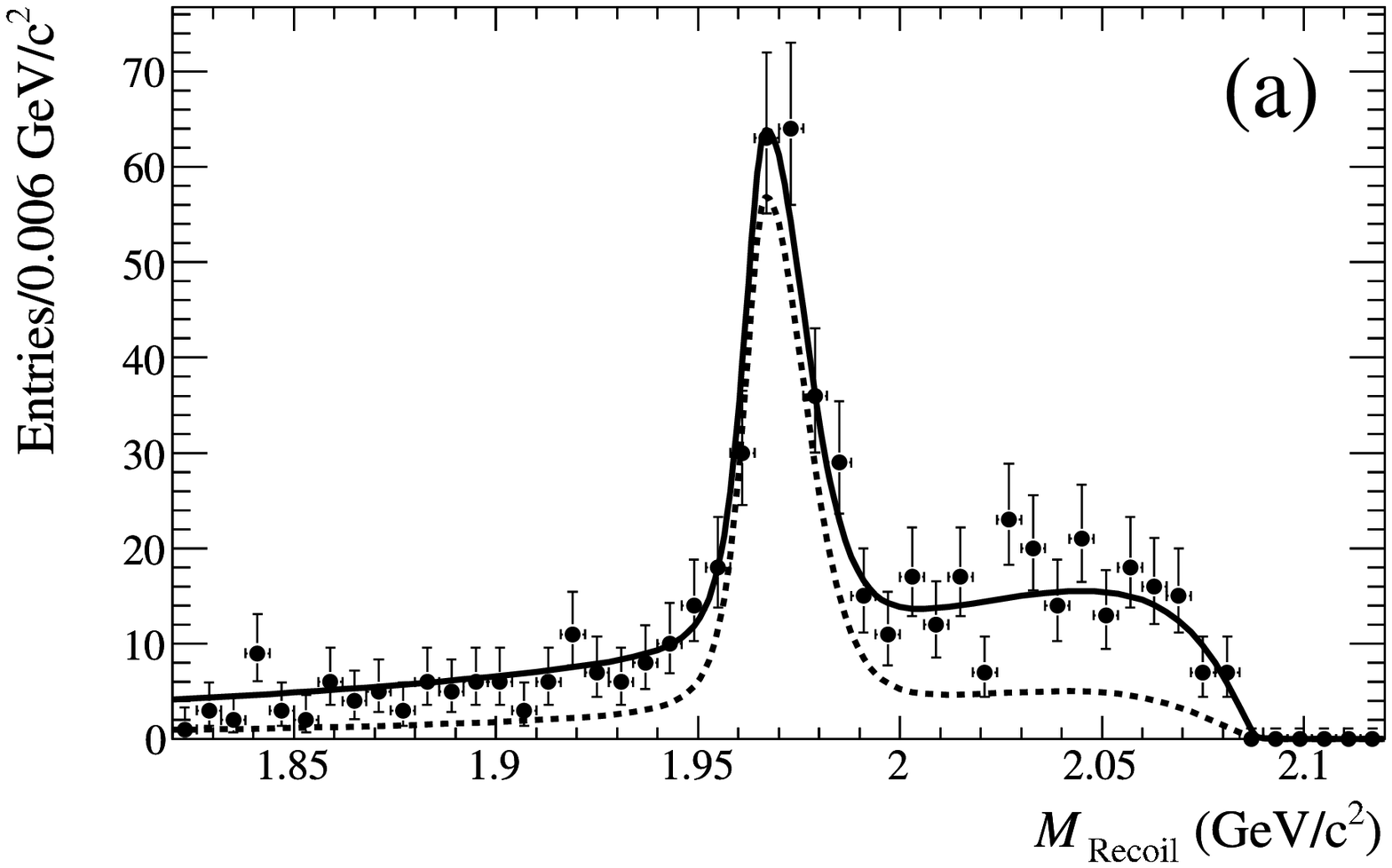} 
  \caption{
    Left two plots show the recoil mass for $\Ds\rightarrow\taup\nut$ ($\taup\rightarrow\ep\nue\nutb$) with $E_{\mathrm{extra}}$ = 0 and $E_{\mathrm{extra}}\>0$. $E_{\mathrm{extra}}$ is the remaining energy in the calorimeter after the full event reconstruction. The two right plots show the corresponding distributions for $\Ds\rightarrow\KS\Kp$ events.
    The solid curve shows the total fit while the dashed-curve shows the signal component. 
  }
  \label{fig:fDs}
\end{center}
\end{figure}

Signal events are reconstructed in the production processes $\ep\en\rightarrow\ccbar\rightarrow\Dss$ $\Dbar_{\mathrm{TAG}}$ $\Kbar^{0,-}$ $X$, with the subsequent decay $\Dss\rightarrow\Ds\g$. Here, $\Dbar_{\mathrm{TAG}}$ is a fully reconstructed hadronic $\Dbar$ meson decay, required to suppress the large background from non-charm continuum $\qqbar$ pair production; $X$ represents a set of any number of pions ($\piz$ and $\pipm$) produced in the $\ccbar$ fragmentation process, and $\Kbar^{0,-}$ represents a single $\Kbar^{0}$ or $\Km$ from $\ccbar$ fragmentation required to balance strangeness in the event. In addition we require a reconstructed \ep which tags the decay $\taup\rightarrow\ep\nue\nutb$. In a similar reconstruction we select $\Ds\rightarrow\KS\Kp$ events.

We extract the signal yields using the \Ds candidate mass determined from the 4-momentum recoiling against the $\Dbar_{\mathrm{TAG}} \Kbar^{0,-} X\g$ system. The fit results are shown in Fig.~\ref{fig:fDs}. We compute the branching fraction using the formula,
\begin{equation}
   \frac{\BR(\Ds\rightarrow\taup\nut)}{\BR(\Ds\rightarrow\KS\Kp) } = 
  \frac{\BR(\KS\rightarrow\pip\pim)}{\BR(\taup\rightarrow\ep\nue\nutb)}
  \frac{(N_{S})^{\tau\nut}}{(N_{S})^{\KS\Kp}}
  \frac{\epsilon^{\KS\Kp}}{\epsilon^{\tau\nut}},
\end{equation}
where $N_{S}$ and $\epsilon$ refer to the number of signal events and total efficiency for the $\tau\nu$ and the normalizing decay modes. The values of the $\KS\rightarrow\pip\pim$ and $\taup\rightarrow\ep\nue\nutb$ branching fractions are obtained from ~\citep{ref:pdg2008}.
We find  $B(\Ds\rightarrow\taup\nut)=(4.5\pm0.5\pm0.4\pm0.3)\%$ and use Eq.~\ref{eq:fDs} to compute \fDs=(233$\pm$13$\pm$10$\pm$7)~MeV \citep{ref:fDs}. Here the errors are statistical, systematic, and due to PDG parameter values.

\section{\large \bf \boldmath
Measurement of $\Dz$-$\Dzb$ Mixing using the Ratio of Lifetimes for the Decays
$\Dz \to \Kmpip$ and $\KpKm$}

Mixing in the charm sector has only recently been observed at B-factories.
One manifestation of \Dz-\Dzb mixing is differing \Dz\ decay time distributions for decays to different \CP eigenstates~\citep{ref:Liu:1994ea}.
We present here a measurement of this lifetime difference using a sample of $\Dz$ and $\Dzb$ decays in which the initial flavor of the decaying meson is unknown.

Assuming \CP\ conservation in mixing, the two neutral $D$ mass eigenstates $| D_1 \rangle$ and $| D_2 \rangle$ can be represented as
\begin{equation}
\begin{array}{rcl}
| D_1 \rangle &=& p | \Dz \rangle + q | \Dzb \rangle \\
| D_2 \rangle &=& p | \Dz \rangle - q | \Dzb \rangle \;,
\end{array}
\label{eq:qpdef}
\end{equation}
where $\left|p\right|^2 + \left|q\right|^2 = 1$. 
The rate of $\Dz$-$\Dzb$ mixing can be characterized by the parameters $x \equiv \Delta m/\Gamma$ and $y \equiv \Delta\Gamma/2\Gamma$, where $\Delta m = m_1 - m_2$ and $\Delta \Gamma = \Gamma_1 - \Gamma_2$ are respectively the differences between the mass and width eigenvalues of the states in Eq.~(\ref{eq:qpdef}), and
$\Gamma = (\Gamma_1+\Gamma_2)/2$ is the average width. If either $x$ or $y$ is non-zero, mixing will occur, altering the decay time distribution of $\Dz$ and $\Dzb$ mesons decaying into final states of specific \CP~\citep{ref:pdg2008}. 
In the limit of small mixing, and no \CP violation in mixing or in the interference between mixing and decay, the mean lifetimes of decays to a \CP eigenstate of samples of $\Dz$ ($\tau^{\Dz}_{hh}$) and $\Dzb$ ($\tau^{\Dzb}_{hh}$) mesons, and the mean lifetime of decays to a state of indefinite \CP ($\tauKpi$), can be combined to form the quantity
\begin{equation}
\displaystyle\yCP = \frac{\langle\tauKpi\rangle}{\langle\tauhh\rangle} - 1,
\end{equation}
where $\langle \tauhh \rangle = (\tau^{\Dz}_{hh} + \tau^{\Dzb}_{hh} )/2$.
An analogous expression $\langle \tauKpi \rangle$  holds for the $\Kmpip$ final state.
If $\yCP$ is zero there is no $\Dz$-$\Dzb$ mixing attributable to a width difference, although mixing caused by a mass difference may be present. In the limit of no direct \CP violation, $\yCP = y$. 

\begin{figure}[t]
\begin{center}
\includegraphics[width=.24\linewidth,height=.24\linewidth,clip]{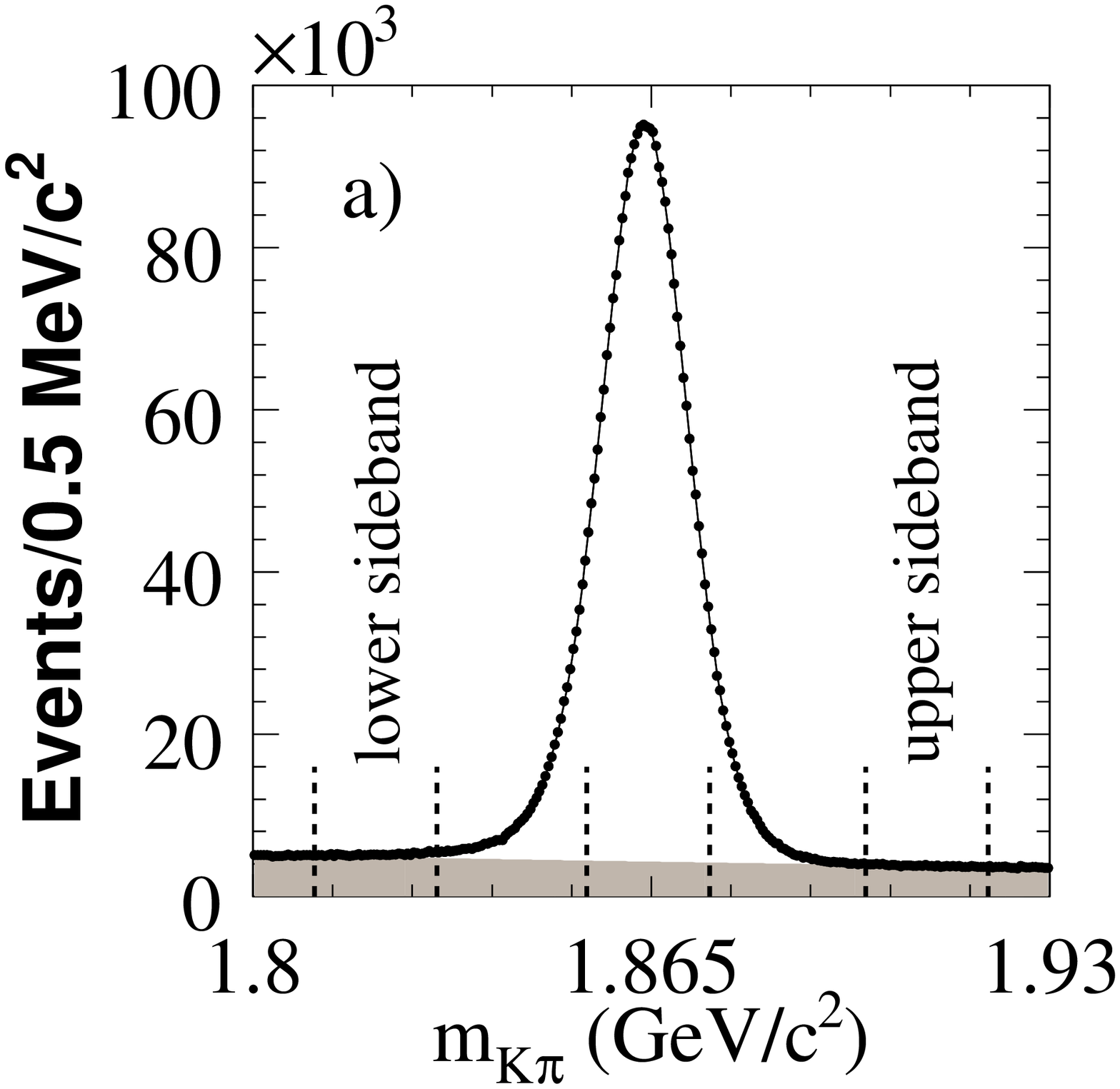}
\includegraphics[width=.24\linewidth,height=.24\linewidth,clip]{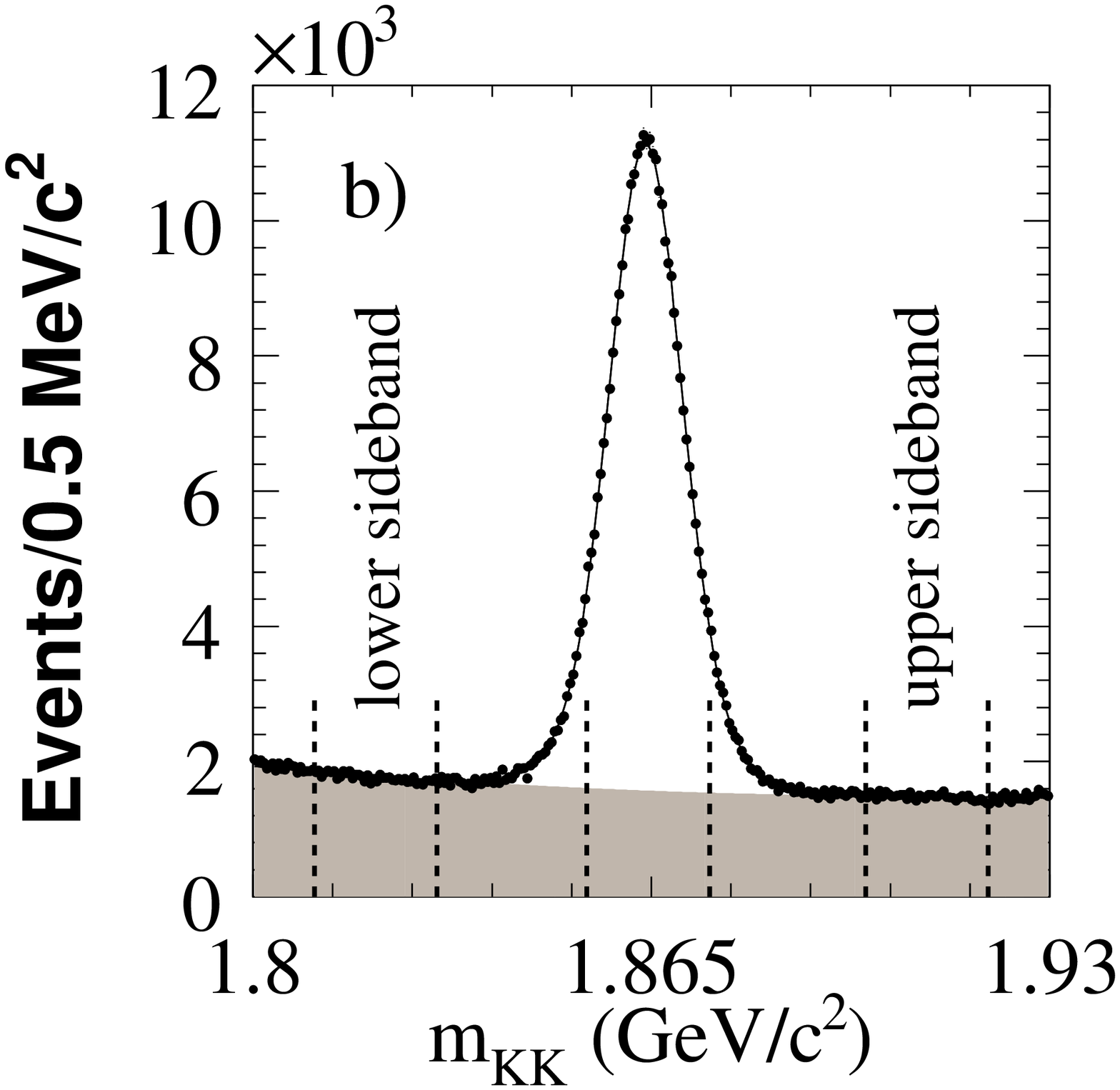}
\includegraphics[width=.24\linewidth,height=.24\linewidth,clip]{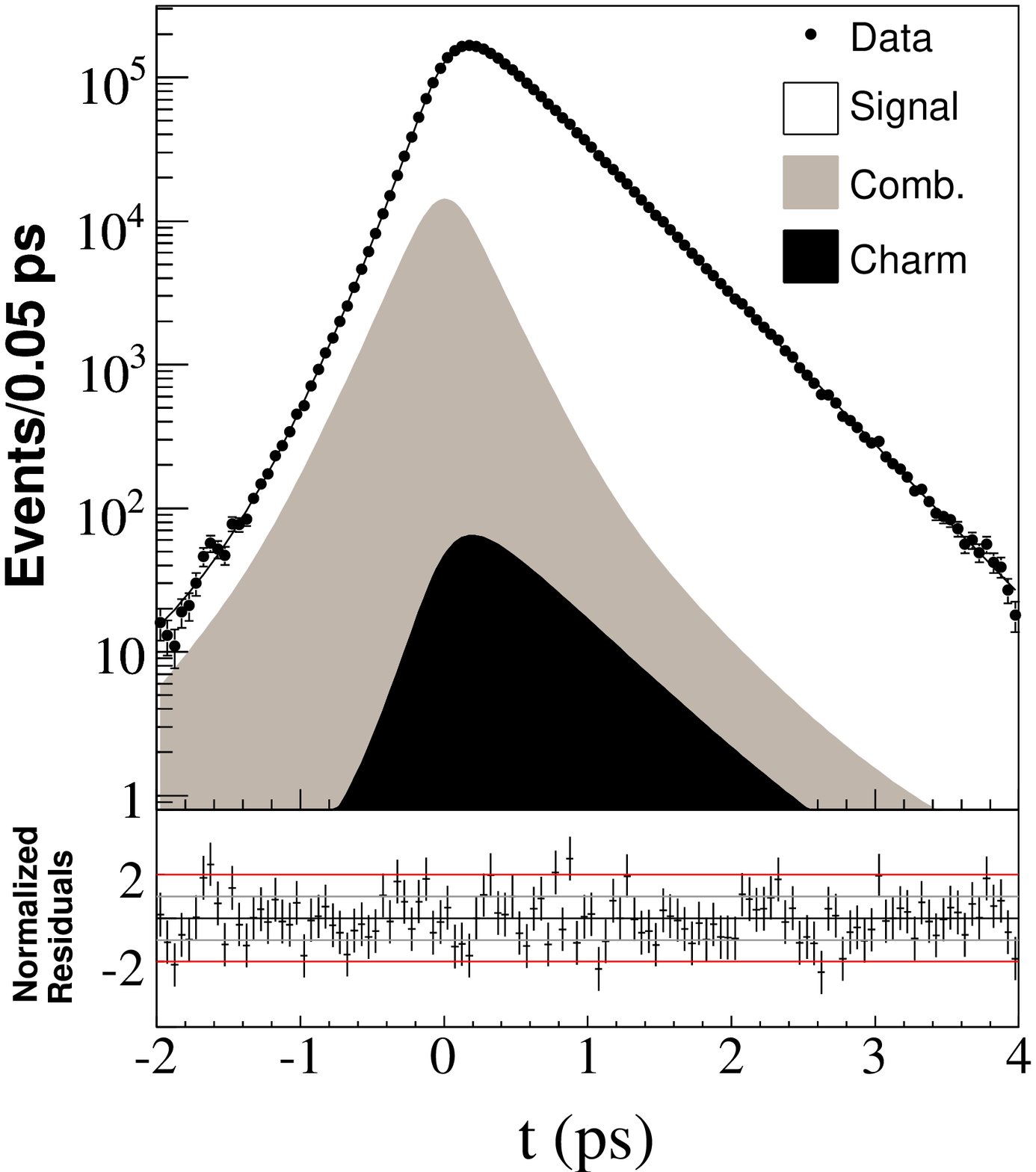}
\includegraphics[width=.24\linewidth,height=.24\linewidth,clip]{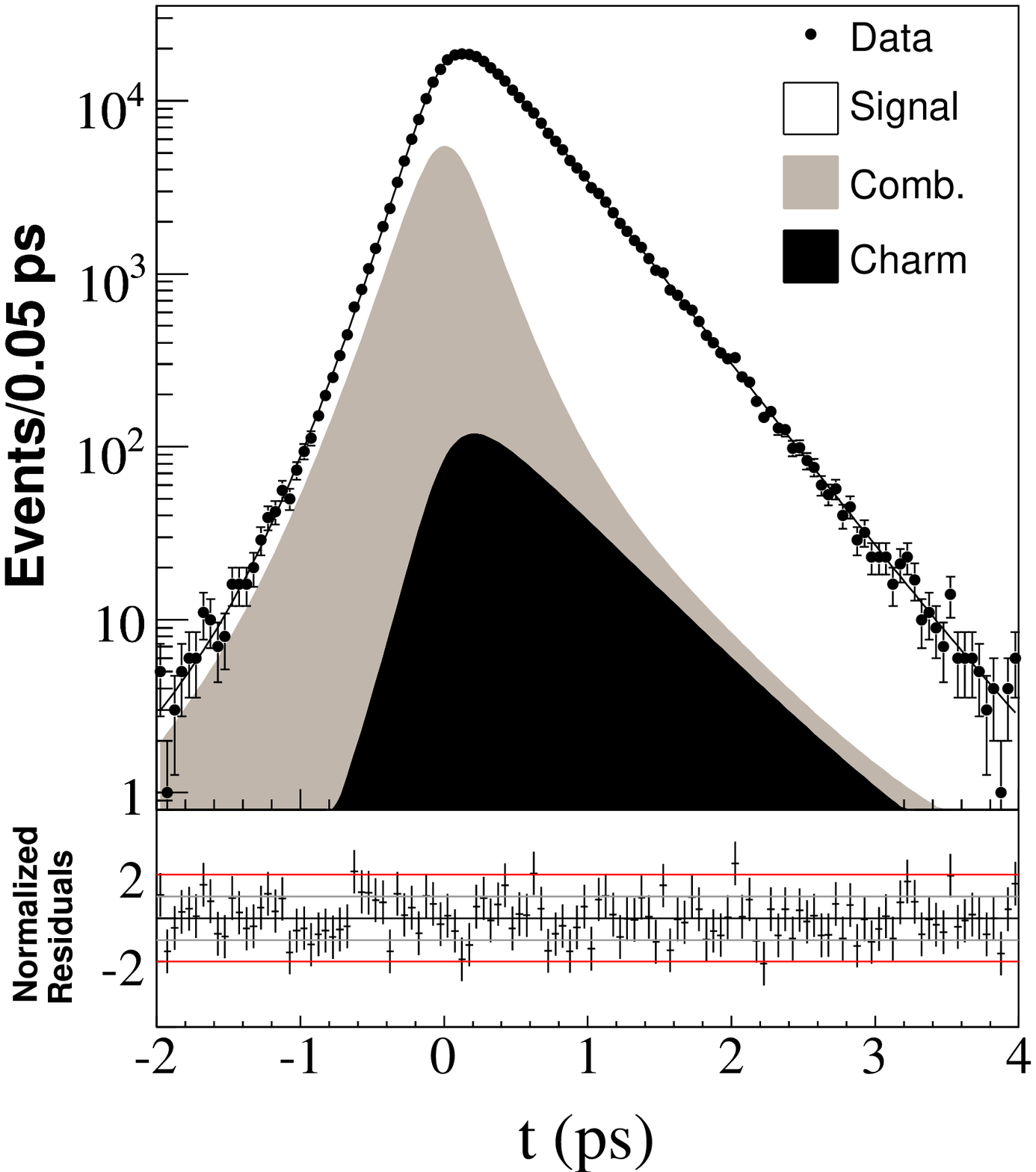}
\caption{
The left plots show the $\Dz$ candidate invariant mass distribution for $\Kmpip$ and $\KpKm$, data are shown by points, total fit as by a curve and background contribution as solid shade. 
The right plots show the decay time distribution for data (points), total lifetime fit (curve), combinatorial background (gray), and charm background (black) contributions overlaid.
}
\label{fig:MassPlot}
\end{center}
\end{figure}

In this analysis we reconstruct \Dz mesons in reactions of the kind $\epem\rightarrow\ccbar\rightarrow\Dz X$ where X is any additional system and \Dz decays to either $\KmKp$ or $\Kmpip$. The decay time of the \Dz candidates is calculated using the measured displacement of the \Dz decay vertex with respect to the \epem interaction region. The \Dz candidate mass distributions and decay time distributions are shown in Fig.~\ref{fig:MassPlot}. To determine the \Dz lifetime in each channel we fit the decay time distributions using an exponential distribution convolved with the resolution function determined from simulated signal events. We find a value of $\yCP({\hbox{untagged}}) = [1.12 \pm 0.26 \stat \pm 0.22 \syst]\%$, which excludes the no-mixing hypothesis at $3.3 \sigma$ \citep{ref:D0}.

\section{\large \bf Search for \CP violation using $T$-odd correlations in $\Dz \to \Kp \Km \pip \pim$ decays}

Physics beyond the SM, often referred to as New Physics (NP), can manifest itself through the production of new particles, probably at high mass, or through rare processes not consistent with SM origins. SM predictions for \CP asymmetries in charm meson decays are generally of $\mathcal{O}(10^{-3})$, i.e. at least one order of magnitude lower than current experimental limits~\citep{ref:chpar}.
Thus, the observation of \CP violation with current sensitivities signal NP. 
We report the results of a search for \CP violation in the decay process $\Dz \to \Kp \Km \pip \pim $ using a kinematic triple product correlation of the form $ C_T = {\bf p_1 \cdot ( {p_2 \times p_3 } ) } $,
where each $ {\bf p_i} $ is a momentum vector of one of the particles in the decay.
The product is odd under time-reversal ($ T $) and, assuming the $ \CPT $ theorem, $ T $-violation is a signal for \CP-violation. 
Strong interaction dynamics can produce a non-zero value of the $ A_{T} $ asymmetries,
\begin{equation}
 A_{T} \equiv \frac{\Gamma(C_T>0) - \Gamma(C_T<0)}{\Gamma(C_T>0) + \Gamma(C_T<0)},
\end{equation}
\noindent 
where $\Gamma$ is the decay rate for the process, even if the weak phases are zero. 
After defining a similar formula for the \CP-conjugate decay process we can construct ${\cal A}_T = \frac{1}{2}(A_T-\overline{A_T})$; a non-zero value of ${\cal A}_T$ would signal CP-violation \citep{ref:bensalem}.

Following the suggestion by I.I.~Bigi~\citep{ref:bigi} to study \CP violation using this technique, the FOCUS collaboration made the first measurements using approximately 800 events and reported $  {\cal A}_T(\Dz \to \Kp \Km \pip \pim) = 0.010 \pm 0.057 \pm 0.037 $~\citep{ref:focus}.
We perform a similar study using approximately $1.5 \times 10^5$ signal events.

\begin{figure}[t]
\begin{center}
\includegraphics[width=.48\linewidth,clip]{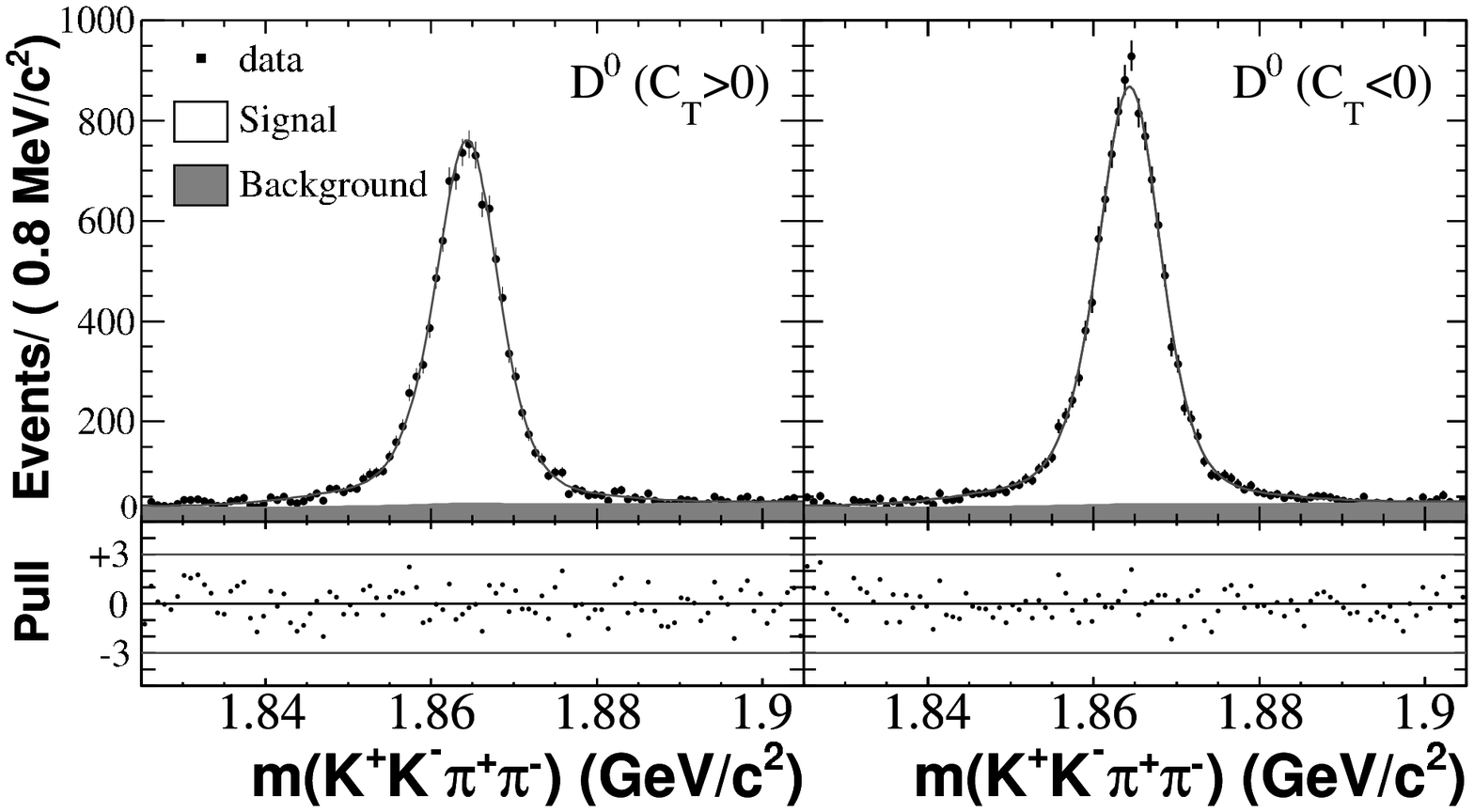}
\includegraphics[width=.48\linewidth,clip]{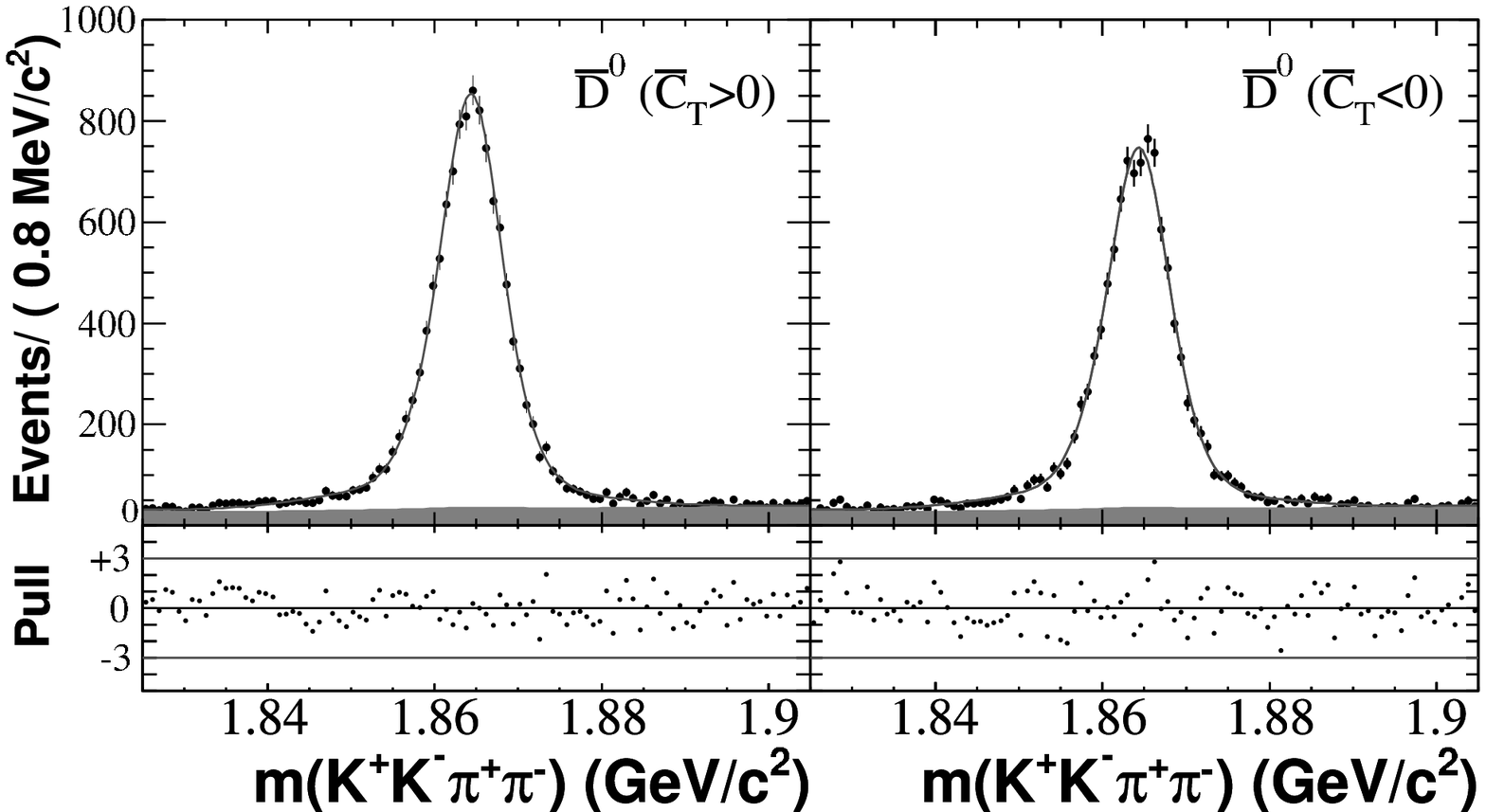}
\caption{Fit projections onto  $m(\Kp \Km \pip \pim)$ for the four different $C_T$ subsamples after a $\Delta m$ signal selection. 
The shaded areas indicate the total backgrounds. The normalized fit residuals, represented by the pulls are also shown under each distribution.}
\label{fig:fig2}
\end{center}
\end{figure}

Reactions of the kind $e^+ e^- \to  X \ \Dstarp; \ \Dstarp\to \pip_s \Dz; \  \Dz \to  \Kp \Km \pip \pim,$ where $X$ indicates any additional (unreconstructed) system, have been selected.
We require the $\Dz$ to have a CM momentum greater than 2.5\gevc. According to the \Dstarp tag and the $C_T$ variable, we divide the total data sample into four subsamples.
The $\Dz$ yields are determined using a binned, extended maximum-likelihood fit to the 2-D (\mKKpipi, \dm) distribution obtained with the two observables \mKKpipi and $\Delta m$ $\equiv$ $ m(\Kp$ $\Km$ $\pip$ $\pim$ $\pisoftp)$ - $m(\Kp \Km \pip \pim)$.
The functional forms of the probability density functions (PDFs) for the signal and background components are based on studies of MC samples. 
We make use of combinations of Gaussian and Johnson SU~\citep{ref:jsu} line shapes for peaking distributions, and we use polynomials and threshold functions for the non-peaking backgrounds. 
Fig.~\ref{fig:fig2} shows the \Kp~\Km~\pip~\pim mass distributions for the four different $C_T$ subsamples. 
The asymmetry determined using the signal yields is found to be consistent with zero: ${\cal A}_T = (1.0 \pm 5.1_{\sta}\pm 4.4_{\sys}) \times 10^{-3}$ with a sensitivity of $\sim0.6\%$ \citep{ref:CPVio}.

\section{\large \bf \boldmath Searches for Lepton Flavor Violation in the Decays $\tau^\pm \rightarrow e^\pm \gamma$ and $\tau^\pm \rightarrow \mu^\pm \gamma$}

Despite the existence of neutrino oscillations~\citep{ref:NuOsc}, decays of \taulg (where $\ell = e, \mu)$ are predicted to have unobservably low rates~\citep{ref:Lee:1977ti} in the SM.
Thus, observation of charged lepton flavor violation would be an unambiguous signature of new physics.
Presently, the most stringent limits are \BRtaueg$<1.1\times10^{-7}$~\citep{ref:Aubert:2005wa} and \BRtaumg$<4.5\times10^{-8}$~\citep{ref:Hayasaka:2007vc} at 90\% C.L., using 232.2\invfb and 535\invfb of \epem annihilation data collected near the \FourS resonance by the \babar\ and Belle experiments, respectively.
This analysis utilizes the entire \babar~ dataset corresponding to a luminosity of 425.5 \invfb, 28.0 \invfb and 13.6 \invfb recorded at the \FourS, \ThreeS and \TwoS resonances, and 44.4 \invfb,  2.6 \invfb and 1.4 \invfb recorded at 40\mev, 30\mev and 30\mev below the resonances, respectively.

The signal is characterized by a $\ell^\pm\gamma$ pair with an invariant mass and total energy in the \CM frame ($E^{\mathrm{CM}}_{\ell\g}$) close to $m_\tau$ = 1.777\gevcc~\citep{ref:pdg2008} and \roots/2, respectively. 
Candidate events must also contain another \mtau decay product (one or three tracks). 
The signal-side hemisphere must contain one photon with \CM\ energy $E^{\mathrm{CM}}_\gamma$ greater than 1 \gev and one track within the calorimeter acceptance with momentum in the \CM\ frame less than $0.77\roots/2$.
This track must be identified as an electron or a muon for the \taueg or \taumg search. 

\begin{figure}[t]
\begin{center}
\includegraphics[width=.4\linewidth,height=.3\linewidth,clip]{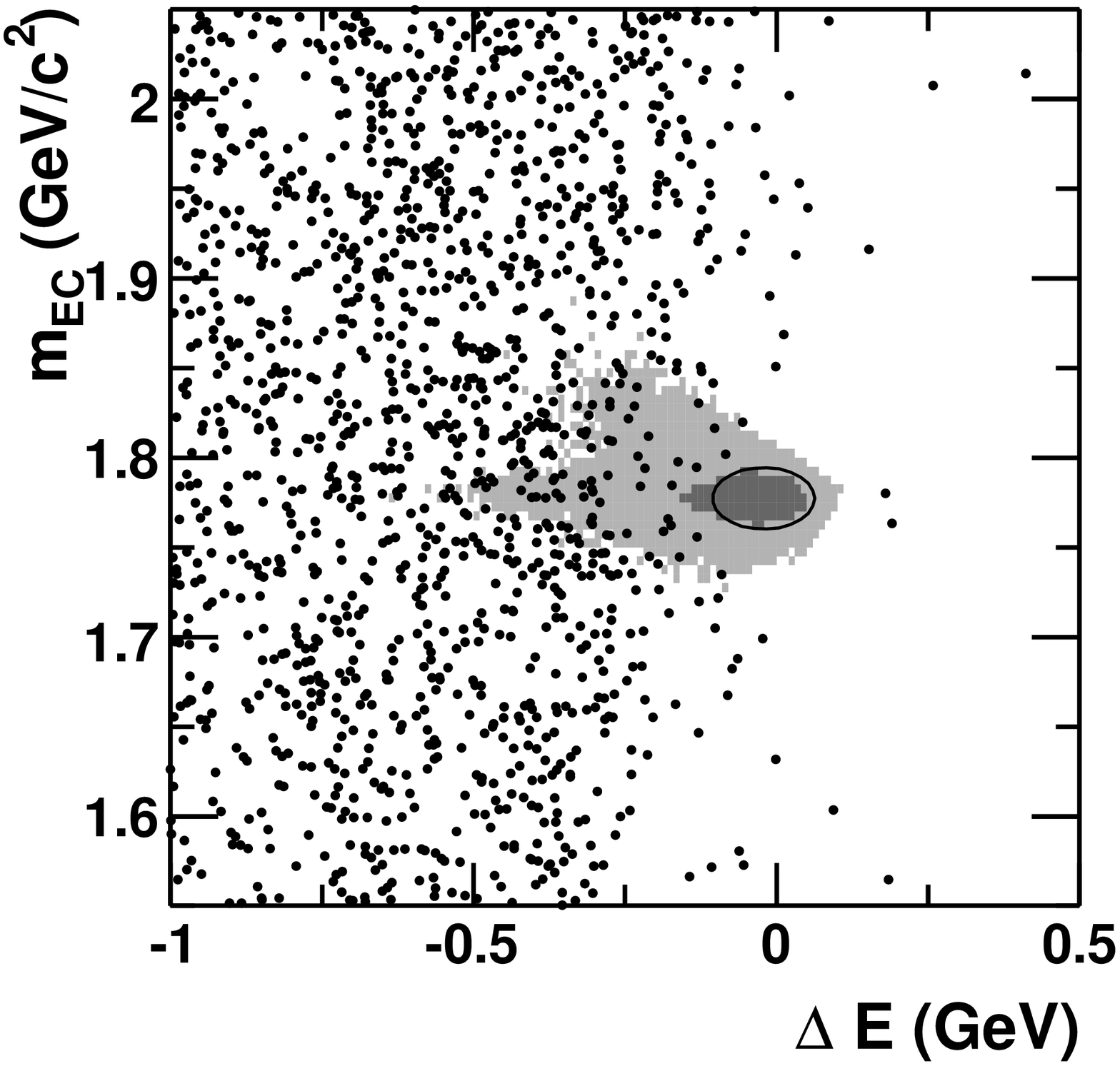}
\includegraphics[width=.4\linewidth,height=.3\linewidth,clip]{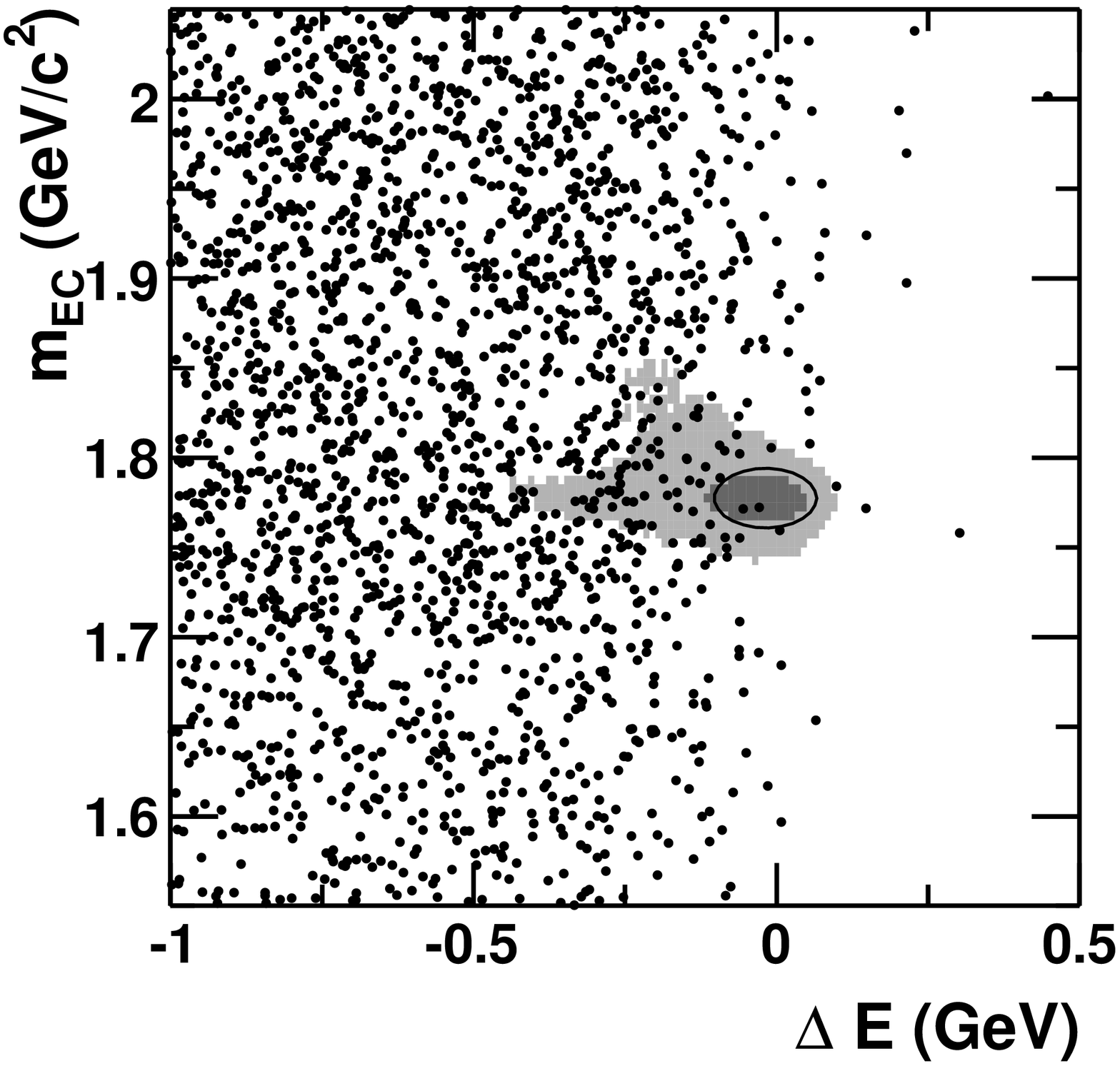}
\end{center}
\caption{
   Distributions of \taueg (left) and \taumg (right) candidate decays in the \mec vs. \DeltaE plane.
   Data are shown as dots and contours containing 90\% (50\%) of signal MC events are shown as light- (dark-) shaded regions. The $2\sigma$ ellipse is shown also.}
\label{ref:fig1}
\end{figure}

Signal decays are searched for using two kinematic variables: 
the energy difference $\DeltaE = E^{\mathrm{CM}}_{\ell\g} - \roots/2$ and  the beam-energy constrained \mtau mass (\mec), obtained from a kinematic fit after requiring the \CM\ \mtau energy to be \roots/2.
For signal events, the \mec and \DeltaE distributions are centered at $m_\tau$ and small negative values, respectively, where the shifts from zero for the latter are due to radiation and photon energy reconstruction effects.
The \mec vs. \DeltaE distributions are modeled by 2-dimensional probability density functions (PDFs) summed over all background event types.
We observe 0 and 2 events for the \taueg and \taumg searches inside the 2$\sigma$ signal ellipse as shown in Fig.~\ref{ref:fig1}.
As there is no evidence for a signal, we set a frequentist upper limit calculated using $\BR^{90}_{UL}=N^{90}_{UL}/(N_{\tau}\eff)$ to be \BRtaueg$<$ 3.3 $\times$ 10$^{-8}$ and \BRtaumg$<$ 4.4 $\times$ 10$^{-8}$ at 90\% C.L. \citep{ref:LepVio}, where \eff is the signal efficiency inside the 2$\sigma$ signal ellipse and $N^{90}_{UL}$ is the 90\% C.L. upper limit on the number of signal events, estimated using the POLE program~\citep{ref:Conrad:2002kn}.

\section{Conclusions}

In conclusion, the \babar~ collaboration continues to exploit its rich data-set to study fundamental aspects of flavor physics. In this article we report an update to our previous measurement of the CKM element $\Vub$ using exclusive \Btopilnu decays. In the charm sector we have extracted a value of \fDs from $\Ds\rightarrow\taup\nut$ decays, we have measured the mixing parameter \yCP using the lifetime ratio $\frac{\langle\tauKpi\rangle}{\langle\tauhh\rangle}$ in \Dz decays, and we have also searched for CP violation using T-odd correlations in 4-body \Dz decays to $\Kp\Km\pip\pim$. Finally, in the tau sector we have placed upper limits on the rates of lepton flavor violating decays $\tau^\pm \rightarrow e^\pm \gamma$ and $\tau^\pm \rightarrow \mu^\pm \gamma$.

\end{document}